\documentclass[prb,twocolumn,nofootinbib]{revtex4-2} 
\usepackage[margin=1in]{geometry}
\usepackage{amsmath}
\usepackage{amsfonts}
\usepackage{xcolor}
\definecolor{bgblue}{RGB}{245,243,253}
\definecolor{darkblue}{rgb}{0, 0, 0.8}
\definecolor{pink}{HTML}{ff00ff} 
\definecolor{darkgreen}{HTML}{00aa00} 
\usepackage{physics}
\usepackage{graphicx}
\usepackage[export]{adjustbox}

\usepackage[skins, theorems]{tcolorbox}
\tcbset{highlight math style={enhanced,
		colframe=white,colback=bgblue,arc=2pt,
		boxsep=4pt, left=5pt,right=5pt,top=2pt,bottom=2pt}
}
\usepackage[hypertexnames=false]{hyperref}
\hypersetup{
	colorlinks   = true, 
	urlcolor     = darkblue,
	linkcolor    = darkblue, 
	citecolor   = darkblue 
}
\usepackage[all]{hypcap}

\newcommand{\figtosymbol}[1]{ 
	{\mathchoice
		{\includegraphics[height=9.6ex, valign=c]{#1}}
		{\includegraphics[height=9.6ex, valign=c]{#1}}
		{\includegraphics[height=7.2ex, valign=c]{#1}}
		{\includegraphics[height=5.4ex, valign=c]{#1}}
	}
}

\newcommand{\smfigtosymbol}[1]{ 
	{\mathchoice
		{\includegraphics[height=8.9ex, valign=c]{#1}} 
		{\includegraphics[height=8.9ex, valign=c]{#1}} 
		{\includegraphics[height=6.6ex, valign=c]{#1}} 
		{\includegraphics[height=5.0ex, valign=c]{#1}} 
	}
}

\newcommand{\ee}{\end{equation}}
\newcommand{\be}{\begin{equation}}
\newcommand{\p}[1]{\left( #1 \right)}
\newcommand{\br}[1]{\left[#1\right]}

\newcommand{\pink}[1]{\textcolor{pink}{#1}} 
\newcommand{\green}[1]{\textcolor{darkgreen}{#1}}

\newcommand{\ca}{\mathcal{A}}
\newcommand{\co}{\mathcal{O}}
\newcommand{\cc}{\mathcal{C}}
\newcommand{\hc}[1]{#1^\dagger} 

\newcommand{\ra}{\rightarrow}
\DeclareMathSymbol{\shortminus}{\mathbin}{AMSa}{"39}
\newcommand{\vac}{\ket{\textrm{vac}}}
\newcommand{\Aout}{\ca_\textrm{out}}
\newcommand{\Ain}{\ca_\textrm{in}}
\newcommand\EatDot[1]{}

\newif\ifSOM
\SOMfalse 

\renewcommand{\thefootnote}{\ifcase\value{footnote}\or$*$\or $\dagger$\or $\ddagger$\or $\mathsection$\or ${*}{*}$\or $\dagger\dagger$\or $\ddagger\ddagger$\or $\mathsection\mathsection$\or  ${*}{*}{*}$\or ${\dagger}{\dagger}{\dagger}$\or ${\ddagger}{\ddagger}{\ddagger}$\or ${\mathsection}{\mathsection}{\mathsection}$\or ${**}{**}$ 
	\fi}

\makeatletter
\def\@fnsymbol#1{\ifcase#1\or a)\or b)\or c) \else\@ctrerr\fi %
}	
\makeatother

\begin{document}
	\setcitestyle{super} 
	
\title{
	{}
	Interference of interference effects
}
	\author{Kevin J.~Randles}
	\email{krandles@uoregon.edu} 
	\author{S.~J.~van Enk}
	\email{svanenk@uoregon.edu}
	\affiliation{Department of Physics and Center for Optical, Molecular and Quantum Science, University of Oregon, Eugene, OR 97403, USA}

\begin{abstract}
	\ifSOM
\else
	{
		We analyze the interference of individual photons in a linear-optical setup comprised of two overlapping Mach--Zehnder interferometers joined via a common beam splitter. We show how two kinds of standard interference effects---namely, single-photon Mach--Zehnder interference and two-photon Hong--Ou--Mandel interference---interfere with one another, partially canceling each other out. This new perspective, along with the overall pedagogical exposition of this work, is intended as an intuitive illustration of why quantum effects can combine nontrivially and, moreover, of the fundamental notion that quantum interference happens at measurement. This work can serve as a bridge to more advanced quantum mechanical concepts. For instance, analyses of this setup in terms of entanglement have a rich history and can be used to test the predictions of quantum mechanics versus local realism (e.g., as in Hardy’s Paradox).\hyperlink{creditAJP}{\textsuperscript{*}}
	} %
	\fi
\end{abstract}
	
	\maketitle
	\stepcounter{footnote}

	\footnotetext[1]{ 
		\raisebox{1.2\ht\strutbox}{\hypertarget{creditAJP}{}}
		A condensed version of  the following article has been accepted by the American Journal of Physics. After it is published, it will be found \href{https://doi.org/10.1119/5.0256745}{here}.
		%
		
	}
	
	\section{Introduction}
	The wave-like behaviors of quantum systems are reflected in the dynamics of so-called \textit{amplitudes}, which are complex numbers associated with the potential outcomes of quantum mechanical processes. Namely, the probability of a particular outcome upon measurement is given by the squared absolute value of the corresponding amplitude.
	The ability to understand, control, and capitalize on the interference of these quantum amplitudes is what much of quantum science is all about. 
	For instance, a main goal of quantum computation is to control the interference of units of quantum information called qubits in order to perform tasks that are either hard for classical computers, interesting (e.g., novel or useful), or ideally both. One realization of quantum computing encodes such qubits in the states of photons,\cite{knill2001scheme,carolan2015universal,kok2007linear,slussarenko2019photonic,bartolucci2023fusion,loqcNote} and related linear-optical networking experiments (e.g., boson sampling \cite{aaronson2011computational,zhong2020quantum}) have been used to demonstrate that quantum computations can achieve things that are (probably\cite{aaronson2013quantum}) hard classically. Moreover, related setups have applications in quantum communication and quantum sensing.\cite{pan2012multiphoton,orieux2016recent,pirandola2018advances}

\begin{figure}[ht!] 
		\includegraphics[width=0.8\linewidth]{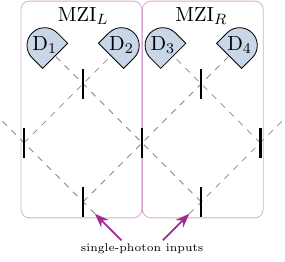} 
		\caption{
			Diagram of the beam-splitter (BS) network setup we consider, which consists of two balanced MZIs  linked by a common central BS. The left and right MZIs (denoted with $L$ and $R$ subscripts, respectively) are both indicated via the bordered regions. All the BSs (shown as vertical black lines) are taken to be 50:50. The incident light is comprised of two identical single photons (shown as arrows), one incident on each of the initial BSs. After traversing the setup (the possible paths are indicated by the dashed lines), each photon will either impinge on one of the four detectors (D$_1 -$D$_4$) or exit the setup via one of the outer BSs (see text for further explanations).
		}
		\label{fig:FullSetup}
	\end{figure}

	In this work, we explore how two quantum interference effects of light, namely, single-photon Mach--Zehnder interference (MZi) and two-photon Hong--Ou--Mandel interference (HOMi), manifest in consonance with one another. (We will review each of these interference effects in Sec.~\ref{sec:generalApproach}.) This exploration naturally leads us to consider the setup depicted in Fig.~\ref{fig:FullSetup} (see Sec.~\ref{sec:fullSetup}). This setup consists of two Mach--Zehnder interferometers (MZIs) linked via a common central beam splitter, forming a network of seven beam splitters that direct incident light along well-defined paths that can intersect one another. We focus on single-photon inputs into each MZI and on a diagrammatic method of analyzing the resulting path interference. 
	
	\subsection{Pedagogical context}\label{sec:pedagogicalContext}
	We note that tutorials, simulations (see App.~\ref{app:exercisesAndSimulation} and Ref.~\citenum{migdal2022visualizing}), and experiments exploring the interference of individual photons can be made quite accessible (e.g., appropriate for a quantum optics or laboratory course for advanced undergraduate or early graduate students),\cite{holbrow2002photon,galvez2005interference,dibrita2023easier,luo2024young} which is important as quantum interference can be a confusing topic to students.\cite{marshman2017investigating,marshman2016interactive,maries2020can,waitzmann2022key} Part of the intrigue of our setup is that, even if one successfully engages with such material and learns about MZi and HOMi, they may be tempted to follow a line of reasoning (similar to that presented in Sec.~\ref{sec:fullSetupNaive}) that is partially quantumly informed, yet nevertheless leads to incorrect conclusions.
	In this work we will highlight several perspectives on where such reasoning goes wrong. 
	
	Our exposition is designed to be accessible and serve as a bridge from the unadorned interference effects that are often introduced to students (MZi and HOMi in our case) to understanding more complicated setups and related fundamental physics,  including photonic quantum computing,\cite{loqcNote} tests of quantum mechanics,\cite{hardy1992quantum,irvine2005realization,lundeen2009experimental} and complementarity.\cite{schwindt1999quantitative,englert2000quantitative,jakob2007complementarity} Throughout the article (see Sec.~\ref{sec:contextAndRelations} especially), we highlight connections between our work and these broader topics, and refer interested readers to the cited references for details. In App.~\ref{app:exercisesAndSimulation}, we provide some resources and exercises that readers can use to further engage with the concepts of this paper and to explore extensions of our setup. For completeness, we provide solutions to these exercises in App.~\ref{app:exerciseSolutions}.
	
	We note that many closely related setups have been analyzed in the literature (starting with Ref.~\citenum{hardy1992quantum}), due to the rich underlying physics. Moreover, the setup we consider has already been thoroughly experimentally investigated in Ref.~\citenum{irvine2005realization}. However, the focuses and intended audiences of these works are quite different than ours (see Sec.~\ref{sec:resolutions} for some additional discussion).
	
	\section{General approach and background}\label{sec:generalApproach}
	We want to analyze and understand how individual photons will propagate through our setup and the resulting probabilities of them being incident on the various detectors placed at possible output ports of the setup (see Fig.~\ref{fig:FullSetup}). 
	In order to focus on the relevant physics and highlight the interference effects we are concerned with, we consider an idealized setup where the photons are confined to a particular finite collection of possible paths (as dictated by the geometry of the setup), the photons are not lost during their propagation through the setup, and the detectors are perfect.\footnote{
		Such issues would undoubtedly be present in a real setup, though they can be largely mitigated via careful experimental design. Moreover, in an actual realization of such a setup, such error mechanisms can (and should) be characterized, so that they can be accounted for when analyzing detection statistics.
	}
	
	\subsection{Underlying principles of quantum mechanics}\label{sec:principlesOfQM}
	To calculate the probability of an initial state transitioning to a certain final state, we will employ the general principles of quantum mechanics outlined by Feynman.
	\cite{feynman1965feynman,feynman1948space} 
	These principles are that (1) the probability of a particular outcome, $\co$, is given by taking the square of the absolute value of a complex number $\ca_\co$ called the probability (transition) amplitude, $P_\co = |\ca_\co|^2$,
	and 
	(2) when $\co$ can occur via multiple indistinguishable processes, the total amplitude $\ca_\co$ is given by adding up the quantum mechanical amplitudes, $\ca_p$, for each such process $p$ as $\ca_\co = \sum_p \ca_p$. Together these imply
	\begin{align} 
		\tcbhighmath[]{P_\co = \bigg| \sum_{p \in p_\co} \ca_p \bigg|^2,} \label{eq:sumOverPaths} 
	\end{align}
	where the sum is over the subset $p_\co$ of all possible processes (or paths) that lead from the initial state of the system to outcome $\co$.\cite{quantumStateNote}
	Here an ``outcome'' corresponds to a possible measurement result of a physical observable such as energy, spin, or---as we consider in this work---the location reached by a photon. Moreover, measuring a red photon at location 1 and a blue photon at location 2 is a different outcome than measuring blue at 1 and red at 2 (see Sec.~\ref{sec:contextAndRelations}). 
	
	This formula encodes the fundamental notions of quantum mechanics, that amplitudes \textit{interfere} and this interference happens at measurement. This can be constructive interference, where the individual amplitudes $\ca_p$, for a given outcome $\co$, have similar phases (as complex numbers) so they build on each other when added up, enhancing $|\ca_\co|$. Alternatively, this can be destructive interference, where the amplitudes have substantially different phases so they tend to cancel each other out when added up, reducing $|\ca_\co|$. The corresponding ability of amplitudes to nicely combine, as opposed to remaining disjoint, is inherited from the linearity of the Schr\"odinger equation. This ``superposition principle''   likewise applies to many waveforms more generally (e.g., those governed by the wave equation or Maxwell's equations in linear media) and leads to a propensity for interference effects in nature.
	
	An important demonstration of such interference is Young's famous double-slit experiment (DSE), wherein light is shone through two small slits in an otherwise opaque slide. The transmitted light is then incident on a screen, on which one finds interference fringes---a series of alternating regions where the light is prominent and absent---which is quite different than the two bright spots one might expect based on superposing the single-slit diffraction patterns of the two lone slits. 
	At its inception,\cite{young1804bakerian} the outcome of the DSE was (classically) well explained by describing light as a wave that goes through both slits and interferes with itself. 
	However, an amazing experimental fact of reality is that the same interference pattern occurs even if the light is sent in one photon at a time.\cite{aspden2016video,luo2024young} Thus, such setups can be used to demonstrate single-photon interference, wherein, for each final location on the screen, the intermediate photon state is in a superposition of having taken one of two separate paths, through one slit or the other. It is the quantum mechanical amplitudes for the photon taking either path that interfere and ultimately (after repeated trials) give rise to the standard double-slit interference pattern.
	
	As we will explore in Sec.~\ref{sec:MZI}, the interference phenomena observed in MZIs can similarly be observed with single photons,\cite{grangier1986experimental,marshman2016interactive,maries2020can} where the two arms of the interferometer take the roles of the two slits in the DSE. Accordingly, although both double-slit interference and MZi can be regarded as phenomena emblematic of the wave-like nature of ``classical'' light, they can each be seen (more fundamentally) as the persistence of a deep-rooted quantum interference effect, wherein the interference happens photon by photon.\cite{electronDSiAndMZiNote} In contrast, the second interference effect we use, HOMi, is not reducible to a classical effect (see Sec.~\ref{sec:HOMint}).

	\subsection{Beam-splitter relations}\label{sec:interferenceAndBSs}	
	A beam splitter (BS) is a linear-optical device that splits light incident on it, letting some transmit and reflecting the rest (up to absorption or scattering loss). 
	In this work, as is standard, we will model the (well-known and characterized) effect that BSs and mirrors have on light propagating through the system without having to worry about the precise nature of their interactions. In particular, the relevant amplitudes, $\ca_p$, can be determined using so-called BS input-output relations, which amount to ``update rules'' for how single-photon amplitudes will change due to a given optical element (BS or mirror).
	
	\begin{figure}[ht] 
		\includegraphics[width=0.75\linewidth]{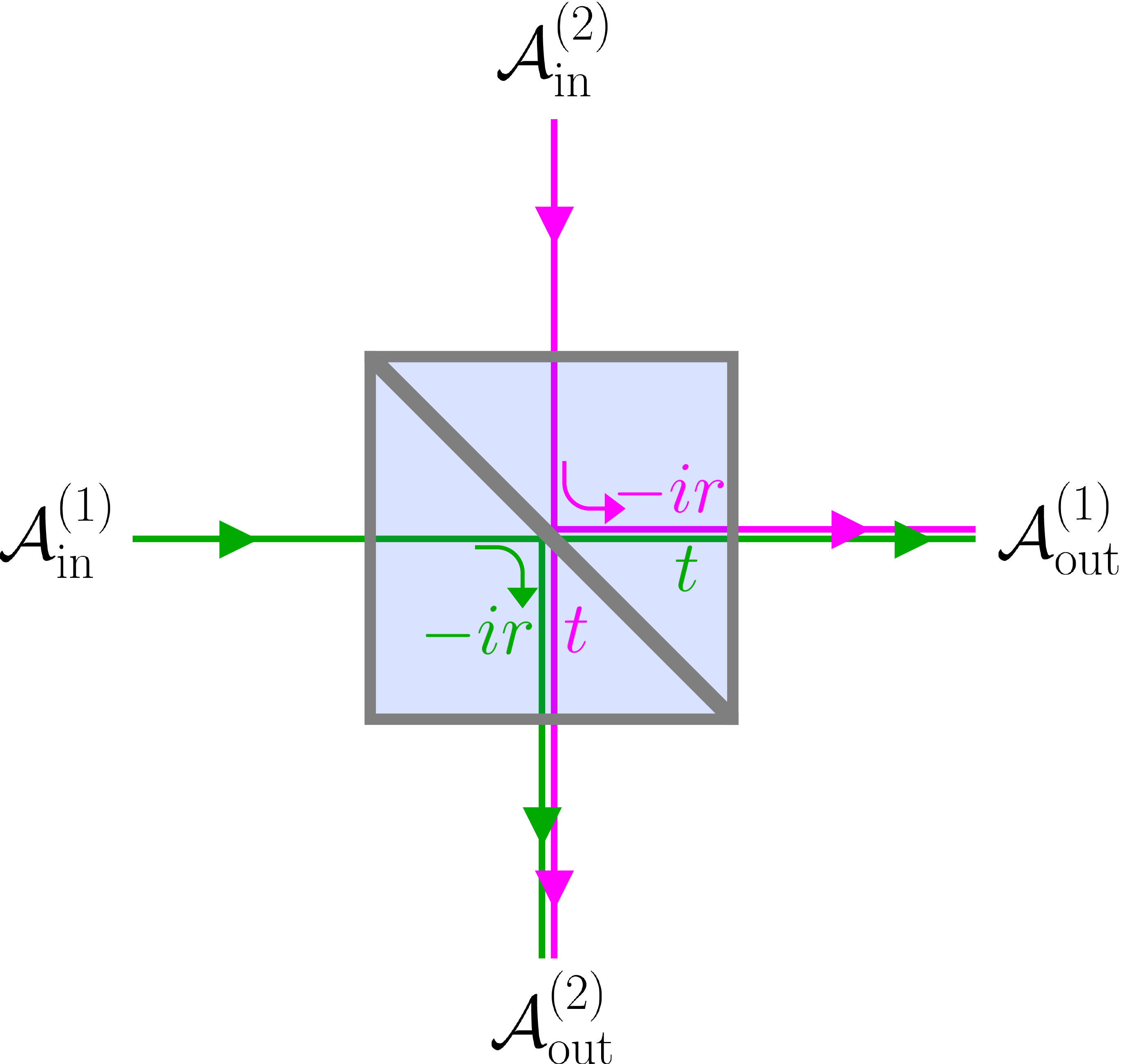} 
		\caption{
			Diagram of a single-photon's interaction with a symmetric BS (see text for further details).
			The input amplitudes $\Ain^{(1)}$ (green) and $\Ain^{(2)}$ (pink) are transformed into the outputs as $\Aout^{(1)} = \green{t} \Ain^{(1)} \pink{- i r} \Ain^{(2)}$ and $\Aout^{(2)} = \pink{t} \Ain^{(2)} \green{- i r} \Ain^{(1)}$. 
			Here we show how the output is formed from the input.
			However, one often knows the initial state and wants to track the output, in which case it is typically more useful to solve for the inputs in terms of the outputs, as in Eq.~(\ref{eq:backPropBS}).
		}
		\label{fig:loneBS}
	\end{figure}
	
	Namely, consider the situation shown in Fig.~\ref{fig:loneBS}, wherein a single photon is in a superposition of being incident on a BS from two directions, 1 and 2, with amplitudes $\Ain^{(1)}$ and $\Ain^{(2)}$, respectively. The interaction with the BS transforms this input photon into an output photon in a new superposition of directions 1 and 2 with respective amplitudes of $\Aout^{(1)}$ and $\Aout^{(2)}$.
	Focusing on the case of symmetric BSs,\footnote{
		 Symmetric BSs are common in optics laboratories, though other phases are possible (for a given $t$ and $r$), 
		 as determined by the optical properties and construction of the BS interface.\cite{hamilton2000phase,symmetricBSNote}
	}
	these single-photon amplitudes are related as
	\begin{subequations}\label{eq:backPropBS}
		\begin{align}
			\Ain^{(1)} &=  t \Aout^{(1)} + i r \Aout^{(2)}, \\
			\Ain^{(2)} &= i r \Aout^{(1)} + t \Aout^{(2)},
		\end{align}
	\end{subequations}
	where $t$ and $r$ are the real transmission and reflection coefficients, respectively, that satisfy $t^2 + r^2 = 1$.
	We will mostly consider 50:50 BSs that equally transmit and reflect light: $t = r = 1/\sqrt{2}$. Note, perhaps more familiarly, that BSs transform electric fields 
	in a manner analogous to Eq.~(\ref{eq:backPropBS}).\footnote{
		More fundamentally, this transformation applies to so-called creation operators yet the corresponding ``second-quantized'' treatment is not needed here nor do we assume the reader is necessarily familiar with it.
		Accordingly, we will present a simpler schematic approach in terms of single photon amplitudes and defer more technical remarks to endnotes.\cite{interferometerUnitaryNote} We refer the curious reader to the Exercise 5 solution, Eqs.~(\ref{eq:distinguishableFinalState})--(\ref{eq:psi23}) in particular, wherein
		we draw the connection between our approach and the second-quantization formalism quite explicitly.
	} 
	
	These relations provide us with an intuitive way to track the phases acquired by light (in the form of individual photons) traveling different paths through a linear-optical setup. Each time light is reflected from a BS or mirror (which is a purely reflective, $r=1$, BS), it acquires a $\pi/2$ phase shift, i.e., the corresponding amplitude is multiplied by $e^{i \pi/2} = i$. Meanwhile, no phase is imparted on transmission through a BS. Thus, 
	the phase factor acquired along a path due to interacting with BSs and mirrors is $i^{N_r}$, where $N_r$ is the number of reflections undergone, and hence 
	light 
	taking different paths to a given destination will interfere. This ``counting method'' allows us to track the phases and magnitudes of each path's contribution \textit{diagrammatically} and thus avoid the 
	underlying matrix formalism while still capturing the fundamental physics.\cite{pathIntegralNote}

	\subsection{Mach--Zehnder interference}\label{sec:MZI}
	Now that we are equipped with the above rules of Feynman and the BS relation of Eq.~(\ref{eq:backPropBS}), we can examine the contributions of different paths that photons could take in propagating from a prescribed starting location to one of the possible outputs.
	We will start by demonstrating 
	the interference of such paths in a Mach--Zehnder interferometer (MZI), which is a fundamental physical instrument, first considered in  Refs.~\citenum{zehnder1891neuer,mach1892ueber}, for measuring phase shifts between two beams with classical and quantum sensing applications. Additionally, the MZI and related setups have offered a platform for testing and exploring some of the fundamental predictions of quantum mechanics.\cite{elitzur1993quantum,busch2006complementarity,pezze2007phase,carolan2015universal} Moreover, the MZI is a key part of the full setup we consider, so it will be useful to identify the underlying physics.
	
\begin{figure}[ht] 
		\includegraphics[width=\linewidth]{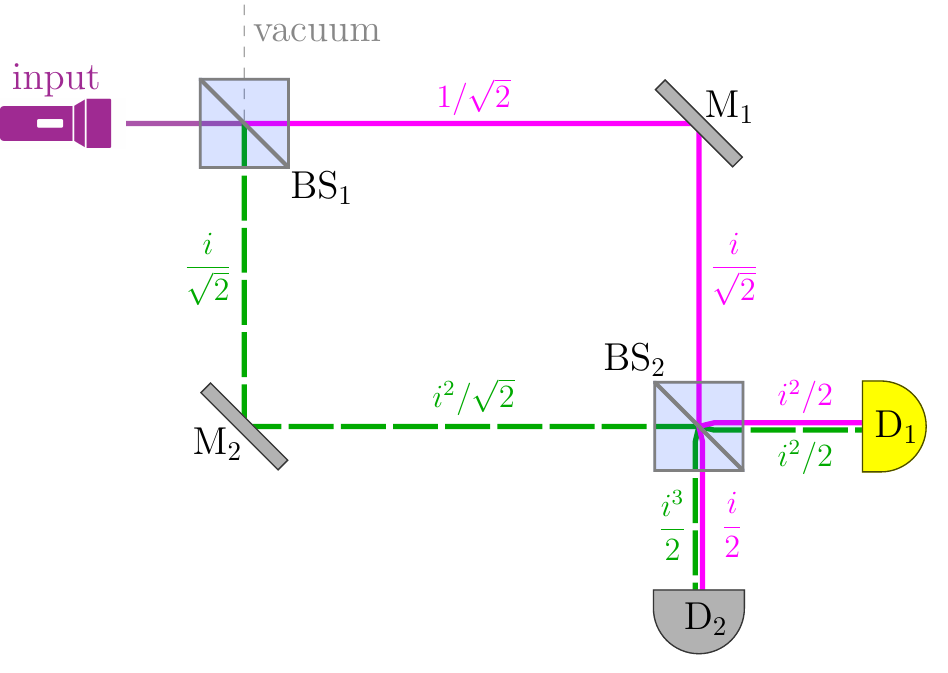}
		\caption{
			Diagram of a balanced MZI with a single input. The incident light can take either the upper (solid pink) or lower (dashed green) path through the interferometer, after which it is directed to one of two detectors, D$_1$ or D$_2$. At each detector, the contributions of these two paths interfere, constructively at D$_1$ and destructively at D$_2$, i.e., the light only impinges on D$_1$ (emphasized in yellow)  and never on D$_2$ (see text for details). Here we use the abbreviations: 50:50 beam splitter (BS), ideal mirror (M), and photodetector (D). 
		}
		\label{fig:labeledMZI}
	\end{figure}

	A typical MZI setup is comprised of a light source, two 50:50 BSs, two mirrors, and two detectors as shown in Fig.~\ref{fig:labeledMZI}.\cite{zetie2000does,holbrow2002photon} We consider light incident on the left port of BS$_1$ from a collimated light source (represented schematically in Fig.~\ref{fig:labeledMZI} by the purple flashlight). Meanwhile, no light is incident on the top port of BS$_1$ (it is in the vacuum state,\cite{vacuumNote} indicated by thin dashing). The incident light can then either transmit through or reflect from BS$_1$ (with equal probability) and then propagate through the upper or lower arm of the interferometer, which are shown as the solid (pink) and dashed (green) paths in Fig.~\ref{fig:labeledMZI}, respectively. Along each path, the light encounters a mirror (M$_1$ or M$_2$) and then the beams are recombined at BS$_2$, from which the light can exit via two possible output ports after which it will impinge on a detector: D$_1$ and D$_2$, for the rightward and downward outputs, respectively.
	
	Focusing on a single-photon input, we will use the counting method introduced above to track the amplitudes acquired along each path through the interferometer (see Ref.~\citenum{grangier1986experimental} for a related experiment). We assume a \textit{balanced} interferometer, wherein light acquires the same net phase during propagation along both arms. This balancing can be accomplished by ensuring that the arms share a common optical-path length, and is often a key preparatory (alignment) step in the operation of such an interferometer and in many optics experiments.
	Importantly, in a balanced setup one can neglect the common phase induced by traversing each arm as it is global, so we only need to track the phases induced by reflections from the BSs and mirrors. (We operate under this assumption throughout the paper.)
	
	We can now determine the amplitudes, $\ca_j$, for the incident photon to reach detector $j=1,2$ as well as the corresponding probability of detection, $P_j = |\ca_j|^2$. For each detector, light can reach it through either the upper or lower arm, so $\ca_j = \ca_{j, \textrm{upper}} + \ca_{j, \textrm{lower}}$. 
	The amplitude of a given path can be found by tracking the contribution acquired due to interacting with all the optical elements (BSs and Ms) encountered along the path. Namely, in the prescribed setup, the BSs split the light 50:50, inducing $1/\sqrt{2}$ factors, the mirrors are perfectly reflective, and a reflection always induces a $\pi/2$ phase corresponding to multiplying by $i$. 
	The two paths that exit BS$_2$ via its bottom output port are both attenuated by two BSs, but the solid (pink) and dashed (green) paths undergo one and three reflections, respectively. Thus, the corresponding amplitude (with path taken being indicated by color) is
	\be\label{eq:MZIA2}
	\ca_2 = \overbrace{\pink{ \frac{1}{\sqrt{2}} \cdot i \cdot \frac{1}{\sqrt{2}} } }^{\textrm{upper path}}
	+ \overbrace{\green{\frac{i} {\sqrt{2}} \cdot i \cdot \frac{i}{\sqrt{2}} }}^{\textrm{lower path}}
	= \frac{\pink{i} + \green{i^3}}{2} 
	= 0,
	\ee 
	so these paths entirely \textit{destructively} interfere. Accordingly, D$_2$ should never detect the incident photon, $P_2 = |\ca_2|^2 = 0$. 
	Meanwhile, the two paths that exit BS$_2$ to the right undergo the same number of reflections (namely two), acquiring a common phase of $\pi$, and thus interfere entirely \textit{constructively}: 
	\be\label{eq:MZIA1}
	\ca_1 = \overbrace{\pink{ \frac{1}{\sqrt{2}} \cdot i \cdot \frac{i}{\sqrt{2}} }}^{\textrm{upper path}} 
	+ \overbrace{\green{\frac{i} {\sqrt{2}} \cdot i \cdot \frac{1}{\sqrt{2}} }}^{\textrm{lower path}}
	= \frac{\pink{i^2} + \green{i^2}}{2}
	= -1,
	\ee 
	so D$_1$ will always detect the incident photon, $P_1 = |\ca_1|^2 = 1$. This Mach--Zehnder interference (MZi) is shown diagrammatically in Fig.~\ref{fig:labeledMZI}, where next to each MZI path segment we display the corresponding amplitude for light that has traversed up to that point. 
	
	\subsection{Hong--Ou--Mandel interference}\label{sec:HOMint}
	When considering multiple photons, one can find surprising interference effects that do not manifest classically. 
	In particular, we will focus on Hong--Ou--Mandel interference (HOMi),\cite{hong1987measurement} wherein two photons are incident on a 50:50 BS (which we will henceforth represent as a thick vertical black line), one on each input port. For the effect to be most prominent, we take the photons to be indistinguishable at measurement	\cite{distinguishableNote}  	
	(i.e., identical in timing, frequency, polarization, and transverse spatial mode)
	and pure.\cite{purityNote}
 	Such 
 	photon pairs might be obtained from a spontaneous parametric down-conversion apparatus as described in Refs.~\citenum{irvine2005realization,dehlinger2002entangled,carlson2006quantum,dibrita2023easier}, though other synchronized sources, such as quantum dots\cite{senellart2017high} or cavity QED based emitters,\cite{randles2024success} could be used.
	Each photon can either transmit or reflect (acquiring a $\pi/2$ phase), so the possible processes for both photons can be schematically summarized as\cite{stateVsAmplitudeNote}
	\be\label{eq:HOMFullState} 
	\p{\smfigtosymbol{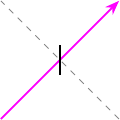} \: + \:  \smfigtosymbol{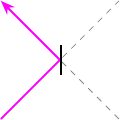}} \!\!
	\p{\smfigtosymbol{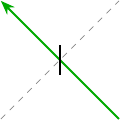} \: + \:  \smfigtosymbol{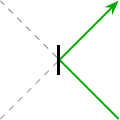}}.
	\ee
	In the shown orientation, the two identical photons propagate up the page as time goes on, ending up either on the left ($L$) or right ($R$). Online, faux coloring is used to schematically keep track of the different processes with pink and green denoting the left and right side inputs, respectively (this is done throughout this subsection and in Fig.~\ref{fig:diagrammaticAmplitudeCalculation}).
	
	The possible outcomes can be organized into three possibilities:
	the photons exit opposite ports (either both transmitting or both reflecting) with amplitude
	\begin{align}\label{eq:HOMLR}
		\ca_{L\&R} &= \figtosymbol{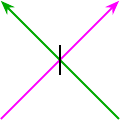} \:  + \: \figtosymbol{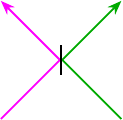} \nonumber \\
		&= \pink{\frac{1}{\sqrt{2}}} \cdot \green{\frac{1}{\sqrt{2}}}
		+ \pink{\frac{i}{\sqrt{2}}} \cdot \green{\frac{i}{\sqrt{2}}}
		= 0
	\end{align}
	(evaluated diagrammatically),
	the photons both go left with amplitude\footnote{
		The additional $\sqrt{2}$ factor present in Eqs.~(\ref{eq:HOMLL}) and (\ref{eq:HOMRR}) is
		necessary for normalization and, more fundamentally, encodes the phenomenon of Bose enhancement (which is implicitly present in our diagrammatic notation whenever there are $n>1$ identical photons in a given mode), i.e., $\p{\hc{a}}^2 \ket{0} = \sqrt{2} \ket{2}$.\cite{boseEnhancementNote} 
		\label{note:BoseEnhancement}
	} 
	\be\label{eq:HOMLL}
	\ca_{L\&L} = \figtosymbol{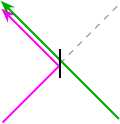}
	= \sqrt{2} \p{\pink{\frac{i}{\sqrt{2}}} \cdot \green{\frac{1}{\sqrt{2}}}}
	=  \frac{i}{\sqrt{2}},
	\ee
	or the photons both go right with amplitude 
	\be\label{eq:HOMRR}
	\ca_{R\&R} = \figtosymbol{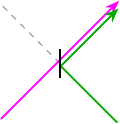}
	= \sqrt{2} \p{\pink{\frac{1}{\sqrt{2}}} \cdot \green{\frac{i}{\sqrt{2}}}}
	=  \frac{i}{\sqrt{2}}.
	\ee
	The two amplitudes for ending up with one photon in $L$ and one photon in $R$ are equal in magnitude yet differ in phase by $\pi$ (due to the difference of two reflections). Hence, they fully destructively interfere: $\ca_{L\&R} = 0$. Accordingly, if detectors are placed after the BS, there will never be a coincident detection at both outputs ($L$ and $R$), $P_{L\&R} \equiv |\ca_{L\&R}|^2 = 0$. Thus, the two input photons always exit together, either both left or both right with equal probability: $P_{L\&L} = P_{R\&R} = |i/\sqrt{2}|^2 = 1/2$. This phenomenon is often deemed the HOM effect.\cite{ogHOMNote} 
	
	\section{Full setup}\label{sec:fullSetup}
	Here we consider a setup that naturally combines the two interference effects we have discussed (MZi and HOMi). This full setup is comprised of two adjacent balanced MZIs connected via a common central 50:50 BS that allows light from one MZI to couple to the other. We imagine using four detectors D$_1 -$D$_4$ (which are taken to be photon-number resolving\cite{PNRDsNote}) to monitor the output of the setup, as shown in Fig.~\ref{fig:FullSetup}. To make each MZI symmetric we also replace the two other (outer) central mirrors by 50:50 BSs (hence every BS in the setup is 50:50). We consider the case where one photon is in the left input and one otherwise identical photon is in the right so that ostensibly HOMi should take place at the central BS and MZi should take place on both sides. We will focus on answering the seemingly innocuous question: \textit{What is the probability of a coincident detection at both inner detectors, D$_2$ and D$_3$}? We denote this probability as $P_{2,3}$.
	Note that because the previously perfectly reflective central mirrors have been replaced by 50:50 BSs, the photons can be lost out of the setup (via either outer central BS, as indicated by the outermost dotted lines in Fig.~\ref{fig:FullSetup} that do not impinge on a detector). 	
	
	\subsection{MZ interference revisited}\label{sec:revisitMZI}
	We will start by considering single-photon (MZ) interference in this setup.
	Suppose only one photon is sent through the full setup, for concreteness, we will take it to be in the left input. Relative to the MZI considered in Sec.~\ref{sec:MZI}, we have replaced the previously perfectly reflective ($=1r$) mirrors by partially transmissive ones (i.e., $r=1/\sqrt{2}$ central BSs).
	The photon will now either 
	be reflected by one of these partially transmissive mirrors and remain in the left MZI, MZI$_L$, or 
	exit MZI$_L$ by transmitting through one of these mirrors. 
	If the photon remains in MZI$_L$, the amplitudes for reaching the corresponding detectors D$_1$ and D$_2$ can be computed in the same manner as in Eqs.~(\ref{eq:MZIA1}) and (\ref{eq:MZIA2}), respectively. The only change is that the path contributions are each reduced by $r$, so $\ca_2$ remains $0$ and $|\ca_1|$ decreases from $1$ to $r$. 
	Thus, if the photon stays in MZI$_L$, 
	full MZi occurs: the photon will always go to D$_1$, never to D$_2$, $P_2 = |\ca_2|^2 = 0$. 
	Otherwise, the photon will exit MZI$_L$, either crossing into the right MZI 
	or being lost out the leftmost central BS. 
	Likewise, as the setup is left-right symmetric, if a single photon was sent into the right MZI, it would never be detected at the corresponding inner detector D$_3$.

	\subsection{Two-photon input: Semi-naive treatment}\label{sec:fullSetupNaive}
	Now we will consider what happens when two identical single photons are sent into the setup, one on each side, as depicted in Fig.~\ref{fig:FullSetup}. 
	We will start by presenting a series of quantumly-informed\footnote{
		We deem this treatment to be ``semi-naive'' in that it leverages actual quantum interference effects (MZi and HOMi) in its reasoning yet misuses them in consonance with one another. 
	} 
	premises from which one can ostensibly deduce whether $P_{2,3}$ is zero. 
	\begin{enumerate}
		\item[(i)] Based on the previous subsection, it seems that neither photon  should reach the inner detector on its own side due to MZi. That is, the left photon should not arrive at D$_2$ nor should the right photon arrive at D$_3$. 
		\item[(ii)] Thus, ostensibly the only way to get a D$_2$D$_3$ coincident detection event would be for the photons to both cross into the other half of the setup, with the initially left  photon crossing to the right and being detected at D$_3$, and the right photon crossing and being detected at D$_2$.
		\item[(iii)] However, when two identical photons are incident on a 50:50 BS, they will always exit the BS together, either both going left or both going right with equal probability. This is the HOM effect discussed in Sec.~\ref{sec:HOMint}. Accordingly, HOMi seems to prevent such a crossing.
	\end{enumerate}
	Thus, it appears that a coincident detection at D$_2$ and D$_3$ should be impossible (such that $P_{2,3} = 0$) as, ostensibly, the photons cannot reach the inner detectors each from their original sides due to MZi and they cannot cross to the opposite sides due to HOMi.
	
	However, the above qualitative reasoning is ultimately wrong, as, in fact, $P_{2,3} > 0$, which is predicted by quantum mechanics (under a more proper treatment). Thus, we have an apparent contradiction (or ``paradox'') between the above reasoning, which predicts $P_{2,3} = 0$, and quantum mechanics. In the next subsections, we will show that quantum mechanics predicts $P_{2,3} > 0$ and then contrast the two treatments to clarify what went wrong in the above semi-naive reasoning.	
	
	\subsection{Two-photon input: Proper treatment}\label{sec:fullSetupProper}
\begin{figure}[ht] 
		\includegraphics[width=0.75\linewidth]{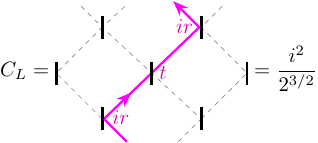}
		\caption{
			Illustrating the counting method for determining the amplitude, $C_L$, of the shown path.
			We annotate the diagram here with the contribution of each transmission and reflection, which are $t$ and $i r$, respectively [see Eq.~(\ref{eq:backPropBS})]. Thus, the path amplitude is $C_L = ir \cdot  t \cdot ir$, which, for the case of all 50:50 BSs, becomes $C_L = i^2/2^{3/2}$ as indicated. The other path amplitudes $O_L$ and $Z_L$ can be calculated similarly (and likewise for side $R$) and are shown in Fig.~\ref{fig:diagrammaticAmplitudeCalculation}. 
		}
		\label{fig:CLDiagAmpExample}
	\end{figure}

	As was done in Eq.~(\ref{eq:HOMFullState}), the amplitude for each possible detection event can be calculated by identifying \textit{all} of the two-photon paths leading to it, applying the counting method to determine the contribution of each path, and then adding up these contributions [in accordance with general principle (2)].\cite{skaar2004quantum,mansuripur2023fundamental} 
	Here we focus on a D$_2$D$_3$ coincidence (though other outcomes can be analyzed similarly). In particular, we want to compute the amplitude, $\ca_{2,3}$, and corresponding probability $P_{2,3} = |\ca_{2,3}|^2$ for this event. The two-photon paths leading to this event can be organized in terms of whether the paths cross or not. 
	Letting $\ca_{S\ra j}$ denote the single-photon amplitude for the photon starting on side $S \in \{L, R\}$ to reach detector $j$, we can thus write the total amplitude for this outcome as
	\be\label{eq:A23SplitInTwo}
		\ca_{2,3} = \ca_{L\ra2} \ca_{R\ra3} + \ca_{L\ra3} \ca_{R\ra2}.
	\ee
	The processes corresponding to $\ca_{L\ra2}$ and $\ca_{R\ra3}$ can happen via either the ``outer path'' (i.e., the corresponding photon transmits, reflects, and then transmits at the subsequent BSs) 
	or the ``zigzag path'' (the photon reflects thrice) 
	with amplitudes  $O_S$ and  $Z_S$, respectively, for starting side $S$. That is, $\ca_{L\ra2} = O_L + Z_L$ and $\ca_{R\ra3} = O_R + Z_R$.
	Meanwhile, the processes corresponding to $\ca_{L\ra3}$ and $\ca_{R\ra2}$ can only happen via the ``crossing path'' (where the photon reflects, transmits, then reflects) 
	with amplitude $C_S$, so  $\ca_{L\ra3} = C_L$ and $\ca_{R\ra2} = C_R$. 
	The corresponding single-photon amplitudes for these paths can be determined via the counting method to be $O_S = i/2^{3/2}$, $Z_S = i^3/2^{3/2}$, and $C_S = i^2/2^{3/2}$ (they are the same for each starting side $S$ 	as our setup is left-right symmetric). 
	These amplitudes can be understood and determined diagrammatically, as demonstrated for $C_L$ in Fig.~\ref{fig:CLDiagAmpExample}.

\begin{figure*}[!ht] 
		\includegraphics[width=\linewidth]{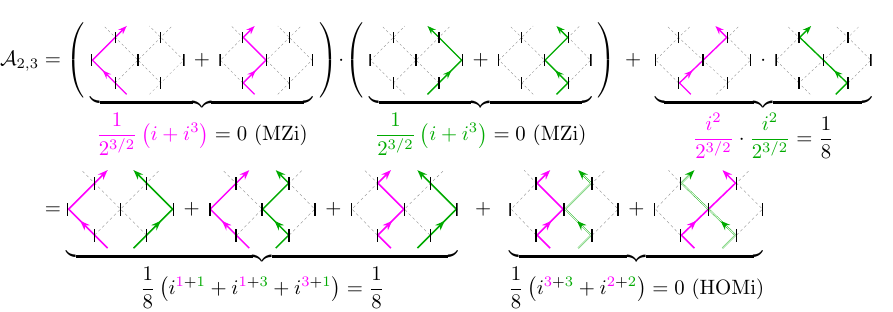}  
		\caption{Diagrammatic calculation of the amplitude $\ca_{2,3}  = 1/8$. Here we visually show how the sum in Eq.~(\ref{eq:A23Expanded}) can be rearranged in two ways to emphasize either MZi (top row) or HOMi (bottom row), as in Eqs.~(\ref{eq:A23MZ}) and (\ref{eq:A23HOM}), respectively. The underbraces highlight how the contributions of different terms can be gleaned by simply counting the number of reflections in a given path. [In the bottom row, we merged products of two single-photon amplitudes into single two-photon amplitudes to emphasize that HOMi is inherently a two-photon interference effect.  Additionally, in the HOMi terms, the $S= R$ (green) paths are hollowed out (making them fainter) 
		to aid in disambiguating the overlapping paths.] 
		}
		\label{fig:diagrammaticAmplitudeCalculation}
	\end{figure*}

\begin{figure*}[ht] 
		\includegraphics[width=0.85\linewidth]{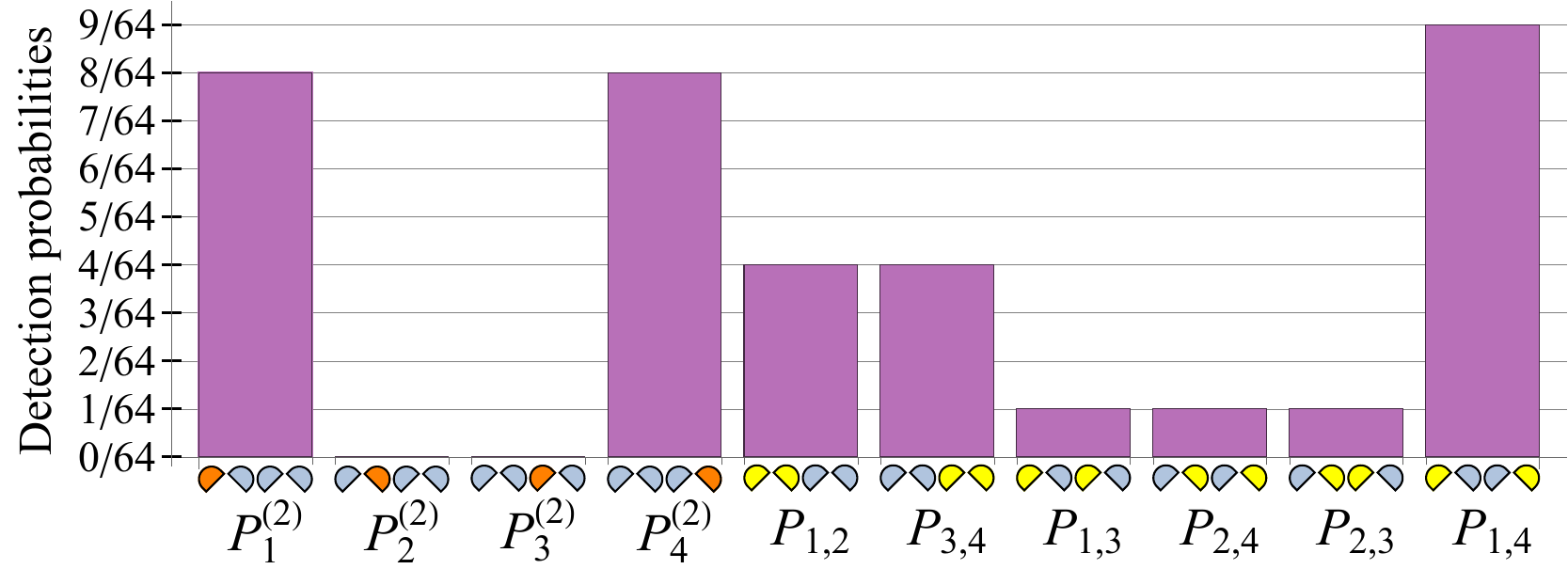} 
		\caption{
			Probability distribution for all possible two-photon detection events in the setup shown in Fig.~\ref{fig:FullSetup}. 
			Here $P_i^{(2)} = P_{i,i}$ is the probability of the two photons (see footnote {\ref{note:BoseEnhancement}})  being incident on the same detector D$_i$ (emphasized in orange) and $P_{i,j}$ is the probability of coincident detection at detectors D$_i$ and D$_j$ (yellow) with $i,j \in \{1,2,3,4\}$. 
			Here we are not showing cases where at least one of the photon exits the setup via of one of the outer central BSs. 
			The probability of losing either individual photon is $|t^2|^2 = 1/4$, so the probability of not losing either photon 
			is $(1 - 1/4)^2 = 36/64 = 56.25 \%$, which is the sum of the probabilities shown above.
		}
		\label{fig:5050setupDetPs}
	\end{figure*}

	Thus, we can rewrite Eq.~(\ref{eq:A23SplitInTwo}) as
	\be\label{eq:A23Expanded}
	\ca_{2,3} = (O_L + Z_L) (O_R + Z_R) + C_L C_R.
	\ee
	We can leverage either MZi ($O_S + Z_S = 0$ for both sides $S$) or HOMi ($Z_L Z_R + C_L C_R = 0$) to simplify Eq.~(\ref{eq:A23Expanded}) as 
	\be\label{eq:A23MZ}
	\ca_{2,3} = C_L C_R 
	= 1/8 
	\ee
	or
	\begin{align}\label{eq:A23HOM}
		\ca_{2,3} = O_L O_R + O_L Z_R + Z_L O_R 
		=   1/8, 
	\end{align}
	respectively. These two reductions are shown diagrammatically in Fig.~\ref{fig:diagrammaticAmplitudeCalculation}. In Eq.~(\ref{eq:A23HOM}), we could leverage MZi to further cancel the $O_L O_R$ term with exactly one of the $O_L Z_R$ or $Z_L O_R$ terms (but not both). Either way, we have that $P_{2,3} = |\ca_{2,3}|^2 = 1/64 > 0$. One can likewise compute the probabilities of the other possible detection events, the results of which are summarized in Fig.~\ref{fig:5050setupDetPs}.
	
	\subsection{Resolutions to apparent contradiction}\label{sec:resolutions} 
	Thus, the qualitative reasoning of the semi-naive treatment, which predicts $P_{2,3} = 0$, contradicts the proper quantum mechanical treatment, which predicts $P_{2,3} = 1/64$. We will now discuss several ways of ``resolving'' this apparent contradiction. Note that $P_{2,3} > 0$ can be and has been confirmed experimentally,\cite{irvine2005realization} so the semi-naive treatment is fundamentally wrong. Accordingly, these resolutions each amount to highlighting a different perspective on where the semi-naive treatment went wrong. 
	
	We focus on three such resolutions, each corresponding to a specific issue in the semi-naive treatment, namely, it wrongly assumes that
	\begin{enumerate}
		\setlength{\itemsep}{1pt}
		\setlength{\parskip}{0pt}
		\item[(1)] MZ and HOM interference happen in isolation, 
		\item[(2)] the two photons are independent, and
		\item[(3)] intermediate probabilities add.
	\end{enumerate}
	The first resolution constitutes our new perspective on this setup and is intended to be intuitive and accessible. The latter resolutions highlight perspectives that have been extensively analyzed in previous works; we include them here to illustrate how issues in the semi-naive treatment can guide one to fundamental quantum mechanical concepts. We elaborate on these resolutions in Sec.~\ref{sec:contextAndRelations}, highlighting their relation to said previous (more technical) works and to each other. Pedagogically, we present the semi-naive treatment---even though it is incorrect---as an example of a flawed line of reasoning that students (even those who know about MZi and HOMi) may be susceptible to, given the challenges often faced when learning about such interference effects (see Sec.~\ref{sec:pedagogicalContext}).\cite{difficultiesNote} Then, the resolutions explicitly address corresponding misunderstandings. 
	
	\subsubsection*{Resolution 1: Interference of interference effects}\label{sec:resolution1} 
	In our setup, MZi and HOMi do not occur independently; rather, a more intricate combination of them occurs.
	This can be seen via the two different reductions of Eq.~(\ref{eq:A23Expanded}) to either Eq.~(\ref{eq:A23MZ}) or Eq.~(\ref{eq:A23HOM}), which are illustrated in the top and bottom rows of the diagrammatic equation in Fig.~\ref{fig:diagrammaticAmplitudeCalculation}, respectively. Mathematically, these reductions are simply an algebraic rearrangement of the complex number $\ca_{2,3} = 1/8$ into a sum of different parts. However, each reduction highlights a different effect: MZi is manifest in Eq.~(\ref{eq:A23MZ}) and HOMi is manifest in Eq.~(\ref{eq:A23HOM}), yet one cannot make both interference effects manifest at the same time. 
	\emph{This is because the zigzag amplitudes $Z$ enter in both effects, and can cancel either the $O$ or $C$ terms but not both.}\footnote{
		This impossibility can be seen by counting: there are five total two-photon paths leading to D$_2$D$_3$ [see Eq.~(\ref{eq:A23Expanded}) or Fig.~\ref{fig:diagrammaticAmplitudeCalculation}], MZi requires four paths (two per side $S$), and HOMi requires two paths, yet $4+2 \neq 5$, so not all amplitudes can be canceled.
	}
	That is, the two interference effects interfere with each other via these shared $Z$ amplitudes.
	
	Accordingly, MZi and HOMi do not happen in isolation in this setup and, moreover, one cannot even say that either effect in particular happens (the physical interpretation of an event should not depend on how we rearrange a sum; see Fig.~\ref{fig:diagrammaticAmplitudeCalculation}). Hence, premises (i) and (iii) in the semi-naive treatment are certainly not valid together and, at the very least, they are individually misleading. Moreover, we did not properly distinguish between there being an amplitude for the photons to traverse a given set of paths and the photons actually traversing said paths. For instance, in premise (iii), after leveraging MZi, it was incorrect to consider two \textit{photons} incident on the center-most BS.  Rather, a single two-photon \textit{amplitude} ($C_L C_R \neq 0$)  describing both photons taking the crossing path remained, as in Eq.~(\ref{eq:A23MZ}).
	
	\subsubsection*{Resolution 2: Photonic entanglement}\label{sec:resolution2}
	In our semi-naive treatment, we implicitly assumed that the two photons behave independently when we reasoned that the probability for a coincident inner detection could be decomposed in a classical way as a sum of two disjoint probabilities: 
	\be\label{eq:snProbAdd}
	P_{2,3} \overset{\textrm{sn}}{=} P_{L\ra2 \& R\ra3} + P_{L\ra3 \& R\ra2}.
	\ee
	Here $P_{L\ra2 \& R\ra3}$ and $P_{L\ra3 \& R\ra2}$ are the respective semi-naive probabilities (indicated via the overset ``sn'') for the photons to either both stay on their initial sides or both cross in a D$_2$D$_3$ coincident detection. According to premises (i)\cite{ifmNote} and (iii), respectively, these probabilities should each be zero, implying $P_{2,3} \overset{\textrm{sn}}{=} 0$. However, this reasoning clearly neglects the possibility of the two photons being entangled, i.e., their joint final state being quantumly correlated. In fact, the presence of nonmaximal entanglement in the photons' state after interacting with the central BSs can be used to account for why $P_{2,3} > 0$ (see Sec.~\ref{sec:contextAndRelations} for additional context).\cite{hardy1993nonlocality,irvine2005realization}
	Thus, such entanglement invalidates the semi-naive reasoning here. However, more broadly, entanglement serves as a fundamental resource in quantum-information processing.\cite{horodecki2009quantum} 

	\subsubsection*{Resolution 3:  Amplitudes versus probabilities}\label{sec:resolution3}
	When analyzing particles that are quantum-mechanically identical, one needs to be careful when 
	attributing specific behavior to an individual particle. For instance, if a D$_i$D$_j$ coincident detection ($i \neq j$) occurs in our setup, assuming the two photons are in fact identical, one cannot say which photon went to D$_i$ and which went to D$_j$ as these outcomes cannot be distinguished (even in principle): they lead to the same final state. In such a situation,  two different processes contribute to the outcome (a D$_i$D$_j$ coincident detection): the initially left photon going to D$_i$ and the right to D$_j$ or vice versa. However, according to the general principles of Feynman, as summarized in Eq.~(\ref{eq:sumOverPaths}), in such a context, the amplitudes for these alternatives add, 
	\be\label{eq:AijIdentical}
		\ca_{i,j} = \ca_{L\ra i} \ca_{R\ra j} + \ca_{L\ra j} \ca_{R\ra i} \quad (i\neq j)
	\ee
	[as in Eq.~(\ref{eq:A23SplitInTwo})], not the corresponding probabilities, $P_{i,j} \neq P_{L\ra i \& R\ra j} +  P_{L\ra j \& R\ra i}$,
	which invalidates the semi-naive reasoning.\cite{feynman1965feynman,semiAstuteNote}
	Thus, in effect, the semi-naive reasoning wrongly assumes that identical outcomes can be distinguished. 
	
	\subsection{Further context and resolution relations}\label{sec:contextAndRelations}
	We intentionally chose these issues and resolutions because their further considerations provide valuable insight into fundamental concepts in physics. 
	For instance, the independence of the photons in Resolution 2 can be made quantitative and precise by assuming that the photons are described by a local hidden variable theory (LHVT). 
	For the purposes of this work, we simply note that such a theory predicts $P_{2,3} = 0$, just as our semi-naive treatment did (though the specific reasoning is quite different), which again seemingly contradicts quantum mechanics. This apparent contradiction is deemed ``Hardy's Paradox'' as it arose in a thought experiment of Lucien Hardy regarding a setup analogous to ours, except with an electron and positron in the overlapping MZIs (instead of photons).\cite{hardy1992quantum} 
	
	Much like the situation we present, the ``paradox'' is that a coincident detection at the inner detectors can occur, even though this suggests that both the particle and anti-particle went through their inner arm and should have annihilated, e.g., forming gamma rays (this is analogous to the photon bunching at the center-most BS in our setup). 
	Related setups have had a rich history as potential means of demonstrating Bell's theorem, which (in brief) says that quantum mechanics cannot be consistently described by any LHVT.\cite{bell1964einstein} For instance, a setup nearly identical to the one we consider was analyzed by Hardy\cite{hardy1992quantumoptical} and  later experimentally tested in Ref.~\citenum{irvine2005realization}, serving as a striking demonstration of Bell's theorem. 
	Note that the (nonmaximal) entanglement of the final photonic state is what ultimately makes it possible to demonstrate Bell's theorem using such a setup.\cite{hardy1993nonlocality} A more complete exposition of entanglement and LHVTs (e.g., what they are and their role in the development of quantum mechanics) is left to other work,\cite{BellTheoremNote} 
	 though we note that corresponding Bell and Hardy tests of local realism (LHVTs) versus quantum mechanics can be explored in the undergraduate laboratory. \cite{dehlinger2002entangled, carlson2006quantum}	
	
	Spurred by Resolution 3, one might ask ``What if the photons are distinguishable?'' and thus consider a variant of the setup wherein the photons have a differing degree of freedom by which final measurements could distinguish them (e.g., perhaps one is red and the other is blue).
	Indeed, we ask this question in Exercise \ref{item:distinguishableImpact}  (and  extend it in Exercise \ref{item:partiallyDistinguishablePhotons}). The short answer is that \textit{distinguishability diminishes interference},\footnote{
		 This applies to single- and two-particle interference effects, as considered here. Systems of more than two particles exhibit rich collective interference, e.g., which underlies boson sampling, with a more intricate dependence on indistinguishability.\cite{tichy2014interference,menssen2017distinguishability} %
		 %
	} which connects Resolutions 1 and 3 (see Solutions 3 and 5 
	for details). 
	Thus, the use of indistinguishable photons is critical for the two-photon interference and concomitant interference of interference effects in our setup. 
	In particular, it is the distinguishability of the outcomes of different processes, i.e., the final state, that determines interference (see Sec.~\ref{sec:IoIConclusions}). 
	This interplay between interference and distinguishability is a manifestation of the principle of \textit{complementarity}, that quantum systems can possess properties that are equally real (and hence measurable) but exclusive (i.e., they cannot be observed simultaneously).  In our setup, and multi-port interferometers more generally, this can be expressed as a quantitative tradeoff between interference visibility and distinguishability,\footnote{
		Note that we have focused on particle distinguishability in this article. 
		However, in the above complementarity contexts, distinguishability is typically expressed as a measure of potential ``which-path information,'' which is related to particle  distinguishability in some contexts, yet more generally characterizes the extent to which one can determine which path a particle took through an interferometer or other setup.\cite{galvez2005interference,marshman2016interactive,maries2020can}
	}
	which can be regarded as a manifestation of wave-particle duality (see Refs.~\citenum{schwindt1999quantitative,englert2000quantitative,jakob2007complementarity} and the references therein for details). 

	The conceptual similarity of Resolutions 2 and 3 hints at a similar connection between entanglement and distinguishability (as well as mixedness\cite{yin2008entanglement}). 
	Indeed, many quantum-information-processing protocols for generating and manipulating entanglement (e.g., remote entanglement generation, entanglement swapping, 
	and photonic fusions \cite{bartolucci2023fusion})
	rely on high-quality interference and thus require indistinguishable photons (or other particles).\cite{rohde2006error,pan2012multiphoton,tichy2014interference,lal2022indistinguishable,jones2023distinguishability,randles2024success} 
	In certain cases, a precise relationship between entanglement and indistinguishability can be given, yet there are subtleties in quantifying the entanglement of indistinguishable particles,\cite{lo2016quantum}
	and the general relation between multipartite entanglement and partial distinguishability is an open question (see  Ref.~\citenum{tichy2014interference} for a tutorial on this topic). 
	Moreover, note that our list of issues and resolutions is not exhaustive. 
	For instance, Ref.~\citenum{lundeen2009experimental} experimentally explores a linear-optical analog of Hardy's thought experiment (which is quite different than that of Ref.~\citenum{irvine2005realization}) and they offer another way of resolving Hardy's Paradox via weak measurement. 
	
	\section{Conclusions}\label{sec:IoIConclusions} 
	In this paper, we analyzed a linear-optical setup comprised of two overlapping Mach--Zehnder interferometers, showcasing how single-photon Mach--Zehnder interference and two-photon Hong--Ou--Mandel interference interplay. We provided a semi-naive line of reasoning to illustrate how someone who knows about these two interference effects separately could easily misapply them in combination with one another and wrongly predict that a certain coincident detection outcome is impossible. We then showed how the possibility of this outcome can be understood as resulting from the interference of these two interference effects (as shown visually in Fig.~\ref{fig:diagrammaticAmplitudeCalculation}). This new perspective is intended to be quite accessible, yet (as we briefly considered) is connected to deeper perspectives and topics explored in the literature such as the Bell's theorem.
	
	Central to these various perspectives (i.e., the resolutions of Sec.~\ref{sec:resolutions}) is the notion that interference (of quantum amplitudes) does not happen at any specific interaction or part of an experiment. Rather, 
	\begin{center}
		\textit{interference happens upon measurement}. 
	\end{center}
	This is in accordance with the rules of Feynman summarized in Eq.~(\ref{eq:sumOverPaths}) and is why we have emphasized the \textit{final} state of the photons (just before measurement).\cite{difficultiesNote} For instance, interference effects that seem to require indistinguishable photon amplitudes (including MZi and HOMi) can still occur for photons that were fully distinguishable when the interference nominally could have taken place, yet were appropriately made indistinguishable thereafter (but before measurement). The quantum-eraser experiment is a single-photon example of this.\cite{scully1982quantum,schwindt1999quantitative,galvez2005interference} 
	%
	Such phenomena can likewise occur in two-photon interference effects.
	For instance, Refs.~\citenum{schrama1991destructive} and \citenum{raymer2010interference} explore setups in which photons with different timing and color, respectively, interfere (see also Refs.~\citenum{kwiat1992observation,pittman1996can}, and Sec.~III D of Ref.~\citenum{pan2012multiphoton}). 
	 %
	This notion is apparent in our setup, as HOMi does not happen at the center-most BS, nor does MZi happen at each MZI. Such impressions could mislead one to the semi-naive premises of Sec.~\ref{sec:fullSetupNaive}. Rather, aspects of both interference effects are imprinted in the final state amplitudes as dictated by the entire setup.
	
	
	
	\appendix
	\section{Exercises and simulation tools}\label{app:exercisesAndSimulation}
	We encourage readers to actively engage with the ideas presented in this work. To facilitate this, here we provide several ``do-it-yourself''-type exercises that are intended to beget further open-ended exploration of our setup. These exercises are intended to be solved through a combination of pen-and-paper calculations and simulation work using the Virtual Lab (VL) developed by Quantum Flytrap (QF).\footnote{
		The Virtual Lab is an excellent resource for simulating photonic interference, including our setup. It can be accessed online at \href{https://lab.quantumflytrap.com/lab}{lab.quantumflytrap.com/lab}, which includes descriptions of how to use it. See Ref.~\citenum{migdal2022visualizing} for further details.
	} 
	Accordingly, each exercise is tagged with at least one of the categories: analytical (A) or Quantum Flytrap (QF). Solutions are provided in App.~\ref{app:exerciseSolutions} and extend beyond what we expect a typical reader would find (e.g., in Exercise 2, we present a general class of solutions along with several notable subcases, whereas identifying these subcases alone should be sufficient for the reader).
	
	We chose to frame the appendices as exercises and solutions because we find the solutions interesting, yet we also think that many readers should be well equipped to solve (at least some of) the exercises. Moreover, the exercises and overall content of this paper are well suited for students to explore in an advanced undergraduate or early graduate course in quantum mechanics, quantum optics, or quantum information. Thus, we encourage instructors of such courses to use these exercises or variations thereof in their teaching (with appropriate acknowledgment).
	
	\begin{enumerate}
		\item \textbf{Simulate the full setup.} \label{item:simulationQFVL} [QF]  
		\\
		Simulate our full setup (shown in Fig.~\ref{fig:FullSetup}) in the QF VL. Suggested workflow:
		\begin{enumerate}
			\item Start by exploring QF's preexisting MZi and HOMi setups as well as the different components available in the VL.
			\item Extend this to the full setup, and run the simulation 
			to see that it is behaving as expected.
			\item Characterize the probability distribution of two-photon detection events and make a plot analogous to Fig.~\ref{fig:5050setupDetPs}.
			Do so by running the simulation $N$ times (determine a reasonable number and modify as necessary) and analyzing the resulting data using your software of choice (a spreadsheet editor is sufficient). 
		\end{enumerate}
		
		\item \textbf{Can we eliminate or enhance the interference of interference effects?} \label{item:adjustingIOIF} [A, QF] 
		\\
		Consider varying the transmission and reflection coefficients of the BSs used in the setup (while preserving left-right symmetry). Assume that none of the path amplitudes are zero so that both MZi and HOMi occur to some extent.
		
		\begin{enumerate}
			\item Can the BS coefficients be altered such that $P_{2,3} = 0$? If so, give an example and discuss whether the semi-naive reasoning is now valid. 
			If not, show why.
			\item How can these coefficients be altered to maximize $P_{2,3}$, while maintaining both full MZi and HOMi?
			[\emph{Hint: The respective conditions for these full interference effects are $O_S + Z_S = 0$ (for both sides $S$) and $Z_L Z_R + C_L C_R = 0$; see Sec.~\ref{sec:fullSetupProper}.}]
		\end{enumerate}
		Explore your findings in QF's VL.
		\item \textbf{What if the photons are distinguishable?} \label{item:distinguishableImpact} [A, QF] 
		\\
		Consider the setup of Fig.~\ref{fig:FullSetup}, except suppose that the photons are fully distinguishable, e.g., they have two different colors or have orthogonal polarizations. How does the two-photon event probability distribution of Fig.~\ref{fig:5050setupDetPs} change? 
		[\emph{Hints: 
			Use the counting method from the main text, but modify it to account for certain outcomes (that were previously indistinguishable) now being distinguishable.
			As in Exercise 1, use QF's VL to explore (the input photons' colors and/or polarizations can be adjusted). Coincident detection probabilities, such as $P_{2,3}$ and $P_{1,4}$, can be shown in the VL (see Fig.~\ref{fig:quantumflytrapscreenshotv2}; the VL's ``beam'' mode can be used to easily show the long-run values of these probabilities).}]			
		\item \textbf{Are single-photon inputs necessary?} [A] 
		\label{item:weakCSs}
		\\
		Rather than using true single-photon inputs, can we see the same ``interference of interference '' phenomenon using weak coherent states as one or both of the inputs? \\
		\emph{Additional context.} To review (or briefly introduce) coherent states, we note that they can be defined by their number state representation (expressed in Dirac notation) as
		\be\label{eq:coherentStateNRep}
		\ket{\alpha} = e^{-|\alpha|^2/2} \sum_{n=0}^\infty \frac{\alpha^n}{\sqrt{n!}} \ket{n},
		\ee
		where $\ket{n}$ is the state with $n$ photons (in the relevant mode) and $\alpha$ is a complex number that characterizes the full state.\footnote{
			In particular, coherent states are subject to Poissonian statistics with the average photon number and variance being equal and given by $\expval{\hc{a} a} = \textrm{Var}(\hc{a} a) = |\alpha|^2$ (see any standard quantum optics textbook for further details\cite{loudon2000quantum,gerry2023introductory}).
		}
		Thus, the corresponding probability of measuring $n$ photons is 
		\be\label{eq:PnCoherent}
		P_n = |\!\braket{n}{\alpha}\!|^2 
		=  e^{-|\alpha|^2} |\alpha|^{2n}/n!
		\ee
		
		Accordingly, for small $|\alpha| \ll 1$, i.e., a ``weak coherent state,'' one can reasonably truncate the sum of Eq.~(\ref{eq:coherentStateNRep}) as higher photon number terms will be suppressed. 
		As we are concerned with two-photon events, we need to keep the terms up to quadratic order ($n=2$): 
		\be
		\ket{\alpha} \approx \frac{\ket{0} + \alpha \ket{1} + \frac{\alpha^2}{\sqrt{2}} \ket{2} }{\sqrt{1 + |\alpha|^2 + |\alpha|^4/2}}.
		\ee
		For example, when $|\alpha| = 0.1$: $P_0 \approx 99.005\%$, $P_1 \approx 0.990\%$, and $P_{n \geq 2} \approx 0.005\%$. This means that when measuring such a state, roughly $99\%$ of the time there will be no photon, yet during the nearly $1\%$ of the time when at least one photon is present, there is a roughly $99.5\%$ chance it will be a single photon. 
		Naturally, this leads to the question: can weak coherent state sources (e.g., weak lasers, which are widely accessible) be considered as effective single-photon sources with efficiency of $P_1$? 		
	
		\item \textbf{What if the photons are partially distinguishable?} \label{item:partiallyDistinguishablePhotons} [A, QF] \\
		Extend Exercise 3 to the case of partially distinguishable input photons, again determining all possible two-photon detection probabilities.
		Does $P_{2,3}$ change? What does this suggest about the ``interference of interference'' phenomenon we discussed? 
		[\emph{Disclaimer: this is more difficult than the previous exercises. 
			Hints: Introduce a parameter quantifying how distinguishable the input photons are (think carefully about how this should be done). If exploring in QF's VL, use a polarization encoding so that distinguishability of the two photons can be nearly continuously varied. Additionally, polarizing BSs can then be used to sort the outputs based on polarization.}]		
	\end{enumerate}

	\section{Exercise solutions}\label{app:exerciseSolutions}
\begin{figure*}[ht!]
		\includegraphics[width=0.9\linewidth]{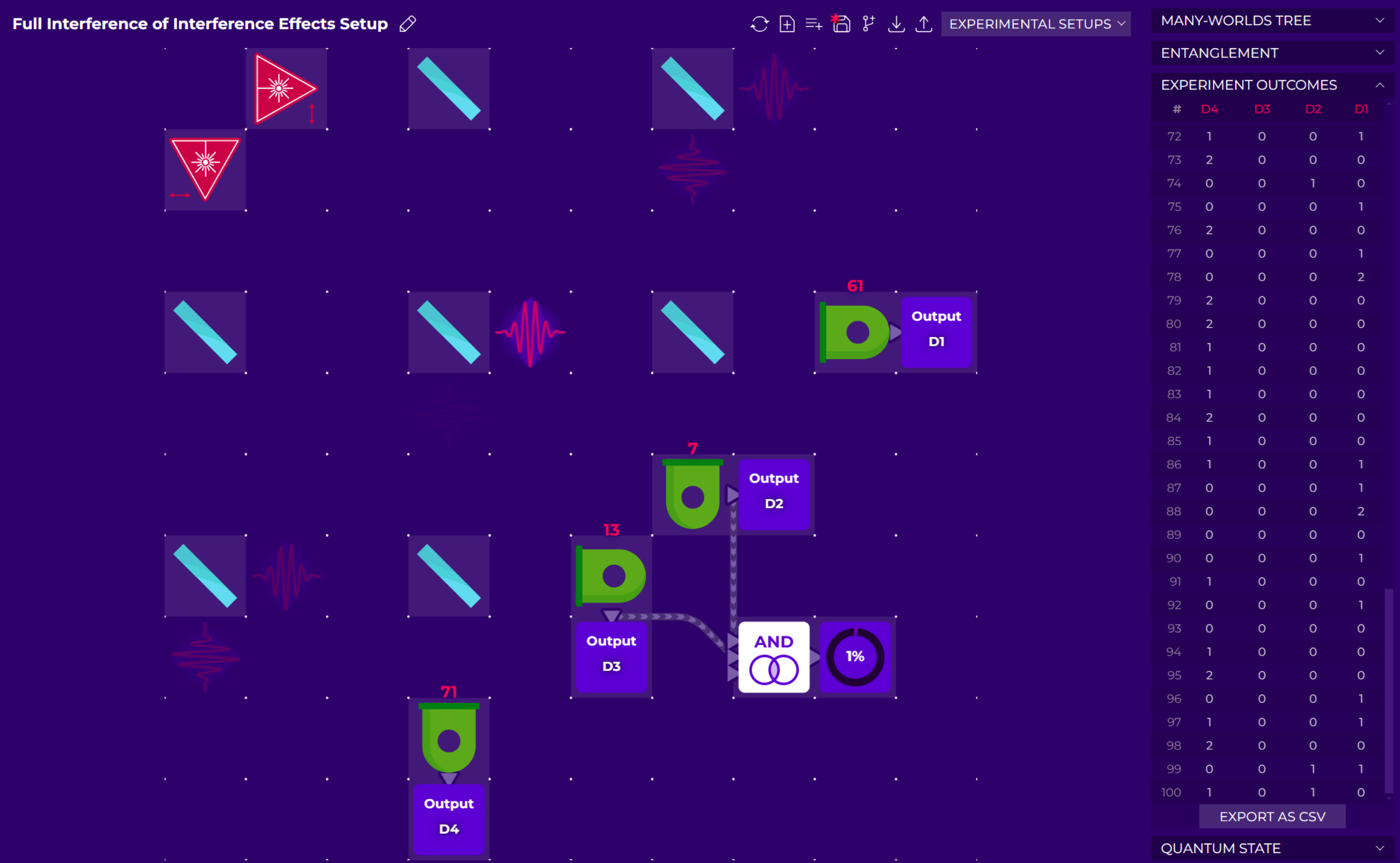}
		\caption{Screenshot of our setup in QF's VL after $100$ experimental trials have been completed. 
			We see that $P_{2,3} \neq 0$ as $1/100 = 1\%$ of the trials have resulted in a D$_2$D$_3$ coincidence (shown using VL's AND logic gate and output ``stat counter'').	
		}
		\label{fig:quantumflytrapscreenshotv2}
	\end{figure*}

\begin{figure*}[ht!]
		\includegraphics[width=0.85\linewidth]{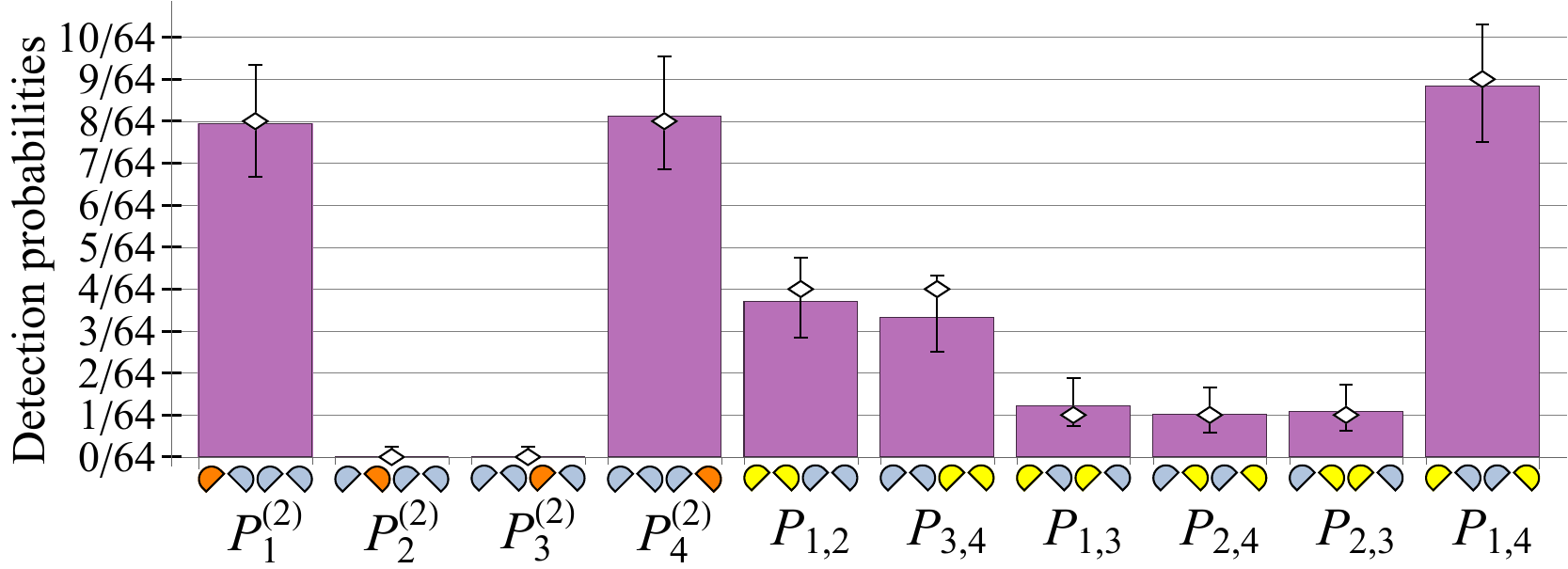}
		\caption{
			Two-photon event probabilities simulated in VL over $N=1000$ trials.
			Error bars indicate the Clopper–Pearson binomial confidence interval with a $95\%$ confidence level, all of which contain the calculated values of Fig.~\ref{fig:5050setupDetPs} (shown as diamonds). 
			[Refer to Fig.~\ref{fig:5050setupDetPs} for variable (and color) descriptions.]
		}
		\label{fig:QFSimulationProbGraph}
	\end{figure*}

	Here we outline potential solutions to the above exercises.
	
	\noindent\\ 1.~\textbf{Full setup simulation.} \\
	Here we show the results of simulating our setup for $N = 1000$ trials in QF's VL. In Fig.~\ref{fig:quantumflytrapscreenshotv2} we show a screenshot of our setup simulated in VL. By analyzing the experimental outcomes of the simulation, we obtain the two-photon event probability distribution shown in Fig.~\ref{fig:QFSimulationProbGraph}, which yields results very similar to what we predicted (see Fig.~\ref{fig:5050setupDetPs}).
	
	\noindent\\ 2.~\textbf{Altering interference of interference.} \\
	(a) We will extend our setup by letting the BS transmission and reflection coefficients ($t$ and $r=\sqrt{1-t^2}$, respectively) vary from the all 50:50 case considered in the main text, while maintaining a left-right symmetric setup.
	In particular, we denote the transmission coefficients of the first, center most, outer middle, and final BSs (in terms of the order of interaction) as $t_0$, $t_c$, $t_m$, and $t_f$, respectively. Then the relevant path amplitudes are
	$O = i t_0 r_m t_f$,
	$Z = i^3 r_0 r_c r_f$,
	and 
	$C = i^2 r_0 t_c r_f$ (we omit the side $S$ labels as the setup is left-right symmetric). To obtain $P_{2,3} = 0$, we need $\ca_{2,3} = 0$, which necessitates that $(O + Z)^2$ cancels $C^2$ by Eq.~(\ref{eq:A23Expanded}).
	It follows that 
	\be
	t_0 r_m t_f - r_0 r_c r_f = \pm r_0 t_c r_f, 
	\ee
	which can be expressed entirely in terms of the reflectivities $R_j \equiv r_j^2 = 1 - t_j^2$ as 
	\begin{align}
		2 R^* &\equiv \frac{R_m (1- R_0) (1- R_f)}{R_0 R_f} \nonumber \\
		&= 1 \pm 2 \sqrt{R_c (1- R_c)} \label{eq:reflectivityConstraint}
	\end{align}
	corresponding to an entire class of setups for which $P_{2,3} = 0$.
	[Note that these solutions are constrained to a circle in the $(R^*, R_c)$ plane of radius $1/2$ centered at $(1/2, 1/2)$.]
	%
	
	To restrict this further, suppose that we want to maintain HOMi in the sense that $Z^2 + C^2 = 0$. 
	Along with the problem assumptions (that $O,Z,C \neq 0$), this constraint corresponds to the $R^* = 1$ case of Eq.~(\ref{eq:reflectivityConstraint}) such that $r_c = t_c = 1/\sqrt{2}$  and $\ca_{2,3} = O (O + 2 Z) = 0$. 
	It follows that $O = -2 Z$, 
	so $Z$ must be suppressed and $O$ enhanced relative to the all 50:50 case wherein $O = -Z$. 
	Several corresponding solutions are highlighted in Table~\ref{table:Reflectivities}. Thus, in order to properly attain the semi-naive value of $P_{2,3} = 0$, one must relinquish either MZi, HOMi, or a combination thereof, altering how the interference effects interfere. Accordingly, the semi-naive reasoning is still not valid; here it wrongly predicts $P_{2,3} > 0$.

	\begin{table}[h]
		\caption{Several reflectivity triplets that satisfy Eq.~(\ref{eq:reflectivityConstraint}) with $(R^*, R_c) = (1, 1/2)$ for which $P_{2,3} = 0$. 
		}
		\label{table:Reflectivities}
		\begin{ruledtabular}
			\begin{tabular}{cccc}
				& $R_m$   & $R_0$          & $R_f$         \\ \hline
				Case 1 & $1/2$   & $1/3$          & $1/3$         \\
				Case 2 & $1$     & $\sqrt{2}-1$   & $\sqrt{2}-1$  \\
				Case 3 & $1$     & $1/2$          & $1/3$        
			\end{tabular}
		\end{ruledtabular}
	\end{table}
	
	\noindent\\ (b) Now, contrary to (a), we want to maximize $P_{2,3}$ by tuning the $t$ and $r$ coefficients of the various BSs. 
	Maintaining both full MZi and HOMi necessitates the relations
	\be\label{eq:MZiApp}
	0 = O + Z = i (t_0 r_m t_f - r_0 r_c r_f)
	\ee 
	and 
	\be\label{eq:HOMiApp}
	0 = Z^2 + C^2 = R_0 R_f (T_c - R_c),
	\ee
	respectively, where we used the amplitudes defined in (a) along with $R_j \equiv r_j^2$ and  $T_j \equiv t_j^2$. 
	
	Equation (\ref{eq:HOMiApp}) implies $T_c = R_c = 1/2$, 
	which we can use along with Eqs.~(\ref{eq:HOMiApp}) and (\ref{eq:MZiApp}) to express $\ca_{2,3}$ in two ways [e.g., see the reductions of Eqs.~(\ref{eq:A23Expanded})--(\ref{eq:A23HOM})]:
	\begin{subequations}
		\begin{align}
			\ca_{2,3} &= C^2 = -Z^2 = \frac{1}{2} R_0 R_f \label{eq:A23CAndZ}\\
			&= -O^2 = R_m T_0  T_f.
		\end{align}
	\end{subequations}
	These expressions for $\ca_{2,3}$ constrain the relationship between $R_0$ and $R_f$ 
	which can self-consistently be used to eliminate $R_f$ (or equivalently $R_0$) as
	\be
	\ca_{2,3} =  \frac{R_0 (1 - R_0) }{R_0/R_m + 2 (1 - R_0)}.
	\ee
	We want to maximize $P_{2,3}$ but, as $\ca_{2,3}$ is real in this case, we can simply maximize it. Using elementary calculus, we find that a maximum of
	$P_{2,3}^{(\textrm{max})} = 17 - 12 \sqrt{2} \approx 2.94 \%$
	occurs for $R_m = 1$ and $R_0 = R_f = 2 - \sqrt{2}$.
	
	This value is roughly $88\%$ larger than the raw value of $P_{2,3} = 1/64 \approx 1.56\%$ in our setup. We note that in Ref.~\citenum{hardy1992quantumoptical}, Hardy considers a version of this setup with $R_c = R_0 = 1/2$, $R_m = 1$, and $R_f  = 2/3$ that nearly saturates the upper bound, achieving $P_{2,3} = 1/36 \approx 2.78\%$ (which is identical to value we would obtain when postselecting on cases where neither photon is lost; see Fig.~\ref{fig:5050setupDetPs}). Part of the intrigue of maximizing this probability is that it makes demonstrating Bell's Theorem (see Sec.~\ref{sec:resolution2}) potentially easier.\cite{hardy1993nonlocality} 
	
	\noindent\\ 3.~\textbf{Distinguishability impact.} \\
	As a concrete example, we take the left and right input photons to be red and blue, respectively, though any other orthogonal modes could be used. 
	The corresponding analysis is simplified for distinguishable inputs (relative to the main text) as identical two-photons effects like Bose enhancement and HOMi will not come into play, and we can talk about which photon does what. 
	In particular, the quantities 
	\be\label{eq:PLiRjDistinguishable}
	P_{L\ra i \& R\ra j} = |\ca_{L\ra i} \ca_{R\ra j}|^2 \quad (i,j \in \{1,2,3,4\}) 
	\ee
	are now well-defined and correspond to the probabilities of the red (left input) and blue (right input) photons reaching detectors $i$ and $j$, respectively. (The amplitudes $\ca_{S\ra j}$ were introduced in Sec.~\ref{sec:fullSetupProper}.) Moreover, the line of reasoning that was semi-naive for identical photons [see Eq.~(\ref{eq:snProbAdd})] is now valid. In particular, as we are not measuring the output colors,\cite{measureColorNote} the coincident detection probabilities 
	are given by a classical-like sum of the corresponding distinguishable outcome probabilities:
	\be\label{eq:PijDistinguishable}
		P_{i,j}^\textrm{(dist)} =
			P_{L\ra i \& R\ra j} +  P_{L\ra j \& R\ra i} \quad (i\neq j).
	\ee
	%
	%
	Comparing this  to the case of identical photons considered in Sec.~\ref{sec:resolution3}, wherein  $P_{i,j}^\textrm{(id)} = |\ca_{i,j}|^2 $ with $\ca_{i,j}$ given by Eq.~(\ref{eq:AijIdentical}), one finds 
	\be 
	P_{i,j}^\textrm{(id)} = P_{i,j}^\textrm{(dist)}  + 
	2 \Re\p{ \ca_{L\ra i} \ca_{R\ra j} \ca_{L\ra j}^* \ca_{R\ra i}^* }
	\ee
	(for $i \neq j$).
	That is, compared to the distinguishable case, $P_{i,j}^\textrm{(id)}$ includes cross terms, which are the signatures of quantum interference.
	Meanwhile, the joint detection probabilities  ($i = j$) are
	\be 
		P_{i,i} = P_{L\ra i \& R\ra i}, 
	\ee
	which may look unchanged but are no longer Bose enhanced (see footnote {\ref{note:BoseEnhancement}}). 
	%
	%

	To evaluate the 16 probabilities, $P_{L\ra i \& R\ra j}$, we start by using the counting method to determine all the relevant single-photons amplitudes:
	\begin{subequations}\label{eq:allSinglePhotonAs}
		\begin{align}
			\ca_{L\ra1} &= \ca_{R\ra4} =  i^2/\sqrt{2}, \\ 
			\ca_{L\ra2} &= \ca_{R\ra3} = 0, \\
			\ca_{L\ra3} &= \ca_{R\ra2} =  i^2/\sqrt{2^3}, \\
			\ca_{L\ra4} &= \ca_{R\ra1} =  i/\sqrt{2^3}.
		\end{align}
	\end{subequations}
	Then the probabilities are given by Eq.~(\ref{eq:PLiRjDistinguishable}), as summarized in Table~\ref{table:distinguishablePij}, from which one can read off the ten color-insensitive two-photon-detection probabilities using Eq.~(\ref{eq:PijDistinguishable}).
	We see that due to a lack of Bose enhancement here, $P_1^{(2)}$ and $P_4^{(2)}$ decrease from their values of $1/8$ in the indistinguishable case to $1/16$. Meanwhile, $P_{1,4}$ increases to $17/64$ from its previous value of $9/64$,
	as the alternatives ($\ca_{L\ra1} \ca_{R\ra4}$ and  $\ca_{L\ra4} \ca_{R1}$) no longer destructively interfere; rather, their corresponding probabilities add. 
	Meanwhile, the other seven two-photon detection probabilities (those involving detectors D$_2$ or D$_3$) are unchanged relative to the main text (see Fig.~\ref{fig:5050setupDetPs}). This is because the amplitude contribution of paths where a photon reaches the inner detector on its own side always vanishes due to MZi ($O+Z = 0$).\cite{semiAstuteNote}
	See the solution to Exercise 5 for a generalization of these results to partially distinguishable photons and for a discussion of the impact of distinguishability on the interference of interference effects phenomenon.
	
	\begin{table}[h]
		\caption{All 16 outcome probabilities $P_{L\ra i \& R\ra j}$ for distinguishable photons (see text for description).}
		\label{table:distinguishablePij}
		\begin{ruledtabular}
			\begin{tabular}{ccccc}
				\multicolumn{1}{c}{} & $L\ra1$ & $L\ra2$ & $L\ra3$ & $L\ra4$  \\
				$R\ra1$ & 1/16 & 0 & 1/64 & 1/64 \\
				$R\ra2$ & 1/16 & 0 & 1/64 & 1/64 \\
				$R\ra3$ & 0 & 0 & 0 & 0 \\
				$R\ra4$ & 1/4 & 0 & 1/16 & 1/16
			\end{tabular}
		\end{ruledtabular}
	\end{table}
	
	\noindent\\ 4.~\textbf{Single photon versus weak coherent state inputs.} \\
	We consider two cases: (1) one weak coherent state input is used and the other input is a genuine single-photon or (2) two weak coherent state inputs are used.
	In case (1), one will most likely get only a single detector click due to the single-photon input and the dominant $n =0$ contribution of the weak coherent state. The next most likely result is the desired one, wherein one obtains a two photon detection event where either two separate detectors click or one detector registers two incident photons (assuming photon-number-resolving detectors for simplicity here). 
	This suggests that our setup (and the predicted interference effects) should work in this case with little modification except for an overall drop in efficiency to about $P_1 \approx |\alpha|^{2}$ for small $\alpha$. 
	Note, however, that the precise event statistics and corresponding error analysis will change.
	
	In case (2), the two sources, which we will call $A$ and $B$, each emit a weak coherent state $\ket{\alpha}$ and $\ket{\beta}$, respectively. We will assume that they have similar mean photon numbers $|\alpha|^2 \approx |\beta|^2$. The issue is that a two photon detection event could be caused by multiple channels:
	(i) the desired one where both sources contribute a single photon with probability 
	$P_{11} = P_1^{(A)} P_1^{(B)} = e^{-|\alpha|^2 - |\beta|^2} |\alpha \beta|^{2}$,\footnote{
			Here we are using the notation $P_{ij} := P_i^{(A)} P_j^{(B)}$ corresponding the probability of a contribution with $i$ photons from source $A$ and $j$ from $B$ (we avoid commas in the subscript here to distinguish these probabilities from those considered in the main text).
	} 
	or
	(ii) the two photons each come from the same arm (which is undesired in that it will lead to a different process and hence different interference than the setup we have considered).
	Note that (ii) can occur in two ways: either two photons from $A$ and none from $B$ contribute, with probability 
	$P_{20} = P_2^{(A)} P_0^{(B)} = e^{-|\alpha|^2 - |\beta|^2} |\alpha|^4/2$,
	or vice versa (none from $A$ and two from $B$), with probability 
	$P_{02} = P_0^{(A)} P_2^{(B)} = e^{-|\alpha|^2 - |\beta|^2} |\beta|^4/2$. For identical sources,  $|\alpha| = |\beta|$, 
	\be
	P_{20} = P_{02} = \frac{1}{2} e^{-2|\alpha|^2} |\alpha|^4
	= \frac{1}{2} P_{11},
	\ee
	so one is equally likely to get one of the undesired outcomes, $P_{20} + P_{02}$, as the desired one, $P_{11}$. [Moreover, one cannot do better by using weak coherent states with different average photon numbers. That is, selecting $|\alpha| = |\beta|$ is the best one can do (in terms of maximizing $P_{11}$ relative to $P_{20} + P_{02}$).]
Thus, although one could indeed use two weak coherent states as the input states to our optical setup, the resulting interference and detection statistics will be very different than case (1) and the case of two single photon inputs (e.g., the output state will not be entangled).

\noindent\\ 5.~\textbf{Partial distinguishability.} \\
\emph{Method background.}
This exercise can be solved purely in terms of amplitudes by combining the counting method introduced in the main text with its adaptation in Exercise \ref{item:distinguishableImpact}.\cite{partialDistNote} 
However, we will instead use this as an opportunity to establish an explicit connection between our approach and the standard second-quantized formalism. The key mathematical tools we will need from this formalism are the so-called creation and annihilation operators $\hc{a}$ and  $a$, respectively. For a given photonic mode, these operators can be defined by their action on photon number states (introduced in Exercise \ref{item:weakCSs}):
$\hc{a} \ket{n} = \sqrt{n+1} \ket{n+1}$
and 
$a \ket{n} = \sqrt{n} \ket{n-1}$ ($n = 0, 1, 2, ...$). That is, $\hc{a}$ creates a photon, while $a$ removes (or ``annihilates'') a photon (hence their names). For a more thorough introduction to these operators and the underlying formalism, see standard quantum optics textbooks.\cite{loudon2000quantum,gerry2023introductory}

\emph{Solution.}
Now consider a case of partially distinguishable input photons described by two orthogonal modes $H$ and $V$ with respective creation operators $\hc{h}$ and $\hc{v}$. 
We take these modes to represent horizontal ($H$) and vertical ($V$) photon polarizations (with the other degrees of freedom of the photons assumed to be fixed and identical). Other orthogonal modes can be considered similarly, e.g., it is common to introduce partial distinguishability via the relative time delay between the two photons.\cite{ogHOMNote,tichy2014interference} In particular, suppose the initial two-photon state is $\ket{\psi_0} = \hc{h}_L \hc{d}_R \vac \equiv \ket{H_L D_R}$, corresponding to left input photon being in the $H$ mode and the right being a ``diagonal'' mode ($D$) defined by 
\be\label{eq:diagCreationOp}
	\hc{d} \equiv \cc \hc{h} + \sqrt{1 - |\cc|^2} \hc{v}.
\ee
Note that here $D$ corresponds to a general superposition of the $H$ and $V$ modes, 
not necessarily an equal superposition with $45^\circ$ polarization. (In the quantum computing nomenclature, one would say that  $\ket{D} = \hc{d} \vac$ represents  a photonic qubit in the polarization encoding.)
The parameter $0 \leq |\cc| \leq 1$ characterizes how distinguishable the two photons are\footnote{
	In particular, the distinguishability parameter $\cc$ is equivalent to the state vector overlap of the two (pure) photons,\cite{randles2024success} i.e., here $\cc = \braket{H}{D}$. One can imagine varying $\cc$ intentionally to see the impact on interference, e.g., using polarization rotators here (as can be explored in QFs VL). However, in practice, the two photons will be partially distinguishable due to experimental imperfections including relative timing errors, mode misalignments, mixedness of the photons,\cite{purityNote} and variability in the photon sources. 
}
and $\vac$ is the vacuum state (corresponding to all modes being in the unoccupied $n=0$ state).
(Operator subscripts of the form $S \in \{L, R\}$ denote the starting side of an input photon and subscripts of $\{1, 2, 3, 4\}$ are later used to denote the final detector reached.)

To translate amplitude-path diagrams (e.g., as used in Fig.~\ref{fig:diagrammaticAmplitudeCalculation}) into a state, we need to account for these distinguishable modes. As in Eq.~(\ref{eq:A23SplitInTwo}), we can split the D$_i$D$_j$ coincident detection outcome into two processes corresponding to paths, where photon $L$ reaches D$_i$ and photon $R$ reaches D$_j$ or vice versa. Crucially, these processes are now partially distinguishable, which must be accounted for in their final state contribution, 
so for $i \neq j$
\begin{align}\label{eq:distinguishableFinalState}
	\ket{\psi_{i,j}} &= \p{ \ca_{L\ra i} \ca_{R\ra j} \hc{h}_i \hc{d}_j + \ca_{L\ra j} \ca_{R\ra i} \hc{h}_j \hc{d}_i } \vac \nonumber \\
	&= \cc \p{
		\ca_{L\ra i} \ca_{R\ra j} + \ca_{L\ra j} \ca_{R\ra i}
	} \ket{H_i H_j} 
	\nonumber \\
	&\quad+ \sqrt{1 - |\cc|^2} \ca_{L\ra i} \ca_{R\ra j} \ket{H_i V_j}
	\nonumber \\
	&\quad+ \sqrt{1 - |\cc|^2} \ca_{L\ra j} \ca_{R\ra i} \ket{V_i H_j}.
\end{align}
We see that for $|\cc| \neq 1$, $\hc{h}_i \hc{d}_j \neq \hc{h}_j \hc{d}_i$, so we cannot factor out a total amplitude as was implicitly done in the main text. Rather, there is an amplitude for each distinguishable outcome $\ket{H_i H_j}$, $\ket{H_i V_j}$, and $\ket{V_i H_j}$. (Note that these amplitudes can be determined by counting, as in the main text, without the need to ever explicitly use creation operators or write down quantum states.) As in Exercise 3, $P_{i,j}$ is a classical-like sum of the probabilities of these distinguishable outcomes, except now the $\ket{H_i H_j}$ outcome is present and admits two-photon interference. 
In particular, as $|\cc|$ decreases from $1$ to $0$ (i.e., the photons are made more distinguishable), it is necessary to gradually shift from adding quantum amplitudes to adding probabilities.\cite{partialDistNote} 

That is, distinguishability destroys interference (see Sec.~\ref{sec:contextAndRelations}).
For $i=j$
\be\label{eq:distinguishableFinalState2}
	\ket{\psi_{i,i}} =  \ca_{L\ra i} \ca_{R\ra i} \hc{h}_i \hc{d}_i \vac 
\ee
with
\be\label{eq:distinguishableFinalState2pt2}
	\hc{h}_i \hc{d}_i \vac =  \cc \sqrt{2} \ket{2H_i} + \sqrt{1 - |\cc|^2} \ket{H_i V_i},
\ee
wherein Bose enhancement accounts for the $\sqrt{2}$ factor in the first term (see footnote \ref{note:BoseEnhancement}, with $\hc{h} \leftrightarrow \hc{a}$) but is not manifest in the second term.\cite{boseEnhancementNote,distinguishableNote} The corresponding single-photon amplitudes can be determined using the counting method and are given in Eq.~(\ref{eq:allSinglePhotonAs}).

We see that for outcomes involving the inner detectors (D$_2$ or D$_3$), one term in the first line of Eq.~(\ref{eq:distinguishableFinalState}) will vanish due to MZi (e.g., as $\ca_{L\ra2} = \ca_{R\ra3} = 0$) and similarly Eq.~(\ref{eq:distinguishableFinalState2}) will vanish. For instance, for a D$_2$D$_3$ detection,
\be\label{eq:psi23}
\tcbhighmath[]{\ket{\psi_{2,3}} = \br{(O+Z)^2 \hc{h}_2 \hc{d}_3 + C^2 \hc{d}_2 \hc{h}_3} \vac,}
\ee
where the $\hc{h}_2 \hc{d}_3$ contribution vanishes due to MZi ($O+Z = 0$) so only the $\hc{d}_2 \hc{h}_3$ term contributes and $P_{2,3} = |C|^4 = 1/64$  remains unchanged. Similarly, $P_2^{(2)}, P_3^{(2)}, P_{1,2}, P_{3,4}, P_{1,3}$ and $P_{2,4}$ remain unchanged relative to Fig.~\ref{fig:5050setupDetPs}.
As seen when solving Exercise 3, only the other three probabilities ($P_1^{(2)}$, $P_4^{(2)}$, and $P_{1,4}$) change. 
%
%
In particular, $P_{1}^{(2)} = P_{4}^{(2)} = \frac{1}{16} \p{ 1 + |\cc|^2 }$, 
so each of these values are reduced by $\frac{1}{16} (1 - |\cc|^2) \equiv \delta P$ relative to the indistinguishable case due to a suppression of Bose enhancement. Correspondingly, $P_{1,4}$ is enhanced by $\frac{1}{8} (1 - |\cc|^2) = 2 \delta P$ to a value of $P_{1,4} = \frac{1}{64} \br{17 - 8|\cc|^2}$ as distinguishability diminishes interference. 
Note that non-destructive measurements\cite{guerlin2007progressive} before the final BSs will similarly diminish interference (as they distinguish different processes and hence provide  `which-path' information). 


\emph{Interpretation.}
From Eq.~(\ref{eq:psi23}), it is easy to see that the coincident inner detection probability $P_{2,3}$ does not depend on whether the photons are identical or not so long as MZi ($O + Z = 0$) occurs.
Someone who does not know about the HOM effect will likely not be surprised by this result: for distinguishable input photons (e.g., one $H$, one $V$), the photons can be reasoned about separately and, as discussed in Solution 3, the semi-naive reasoning correctly predicts the value of $P_{2,3} = |C_L C_R|^2 = 1/64$ [as premises (i) and (ii) hold, while HOMi no longer seems to preclude a crossing in premise (iii)]. Without knowing the HOM effect, this probability need not necessarily vary with photon distinguishability. However, if one does know of the effect, they might (semi-naively) expect that continuously changing the polarization of the $V$ photon to $H$ should make it necessary to take HOMi into account. Then, one explanation for why the HOM effect does not seem to work in this setup, i.e., that $P_{2,3}(\cc)$ is constant, is in terms of interference of interference.

One might be tempted to interpret $P_{2,3}(\cc)$ being constant as an indication that HOMi does not play a role in the processes leading to a D$_2$D$_3$ coincident detection. However, if the final BSs are omitted\cite{hardy1992quantum,irvine2005realization} or the photons are  non-destructively measured before they reach the final BSs, one will indeed see extremely different results depending on how distinguishable the input photons are. For instance, with the final BSs removed, identical photons will never exit opposite ports of the central BS (due to HOMi), whereas distinguishable photons indeed can. In particular, even though $7/10$ of the considered two-photon detection probabilities do not depend on $\cc$, the entanglement structure of the final photonic state does, which has important implications regarding the potential to demonstrate Bell's theorem in a setup\cite{hardy1993nonlocality} (e.g, such a demonstration could not be done with 
distinguishable photons, $\cc = 0$). 

\bibliographystyle{unsrt} 
\bibliography{interferenceOfInterference_SOM_arXiv} 

\begin{thebibliography}{10}
\bibitem{symmetricBSNote}
Provided the relative phases acquired along different paths are appropriately
  tuned and stabilized, the exact BS phases will not change our ultimate
  conclusions and can be treated as a convention.
\bibitem{stateVsAmplitudeNote}
For completeness, note that the pictorial representation of different photon
  paths in Eq.~(\ref{eq:HOMFullState}) is being used to represent the final
  state of the system (a vector in Hilbert space\cite{quantumStateNote}),
  whereas in subsequent expressions, including
  Eqs.~(\ref{eq:HOMLR})--(\ref{eq:HOMRR}), such symbols are used to denote
  amplitudes (which are complex numbers).
\bibitem{PNRDsNote}
One could postselect on coincident detections at different detectors to avoid
  the need for number-resolving detectors.
\bibitem{semiAstuteNote}
The astute reader may note that if one defines these semi-naive
  ``probabilities'' in a different quantumly informed way as $P_{L\ra i \& R\ra
  j} \equiv |\ca_{L\ra i} \ca_{R\ra j}|^2$, then Eq.~(\ref{eq:snProbAdd})
  appears to be valid. Indeed, for $i=2$ and $j=3$, this yields the correct
  probability, $P_{2,3}$, so long as there is full MZi (namely, $\ca_{L\ra2}
  \ca_{R\ra3} = 0$) as then there are no cross terms in
  Eq.~(\ref{eq:sumOverPaths}). However, this approach is not valid as it cannot
  be applied for $i,j \in \{1,4\}$ or, more generally, when full MZi does not
  occur, i.e., so long as both such ``probabilities'' are nonzero. See the
  solutions to Exercises \ref{item:distinguishableImpact} and
  \ref{item:partiallyDistinguishablePhotons} for some related discussion.
\bibitem{partialDistNote}
The probability of each detection event, $\mathcal{E}$, is given by a weighted
  sum of its values in the fully distinguishable and indistinguishable cases,
  denoted here as $P_\textrm{dist}$ and $P_\textrm{indist}$, respectively. In
  particular, after introducing a mode-overlap parameter $\mathcal{C}$ [as in
  Eq.~(\ref{eq:diagCreationOp}), whose squared modulus $|\mathcal{C}|^2$ is the
  probability that the two inputs will be in the same mode], we have
  $P_\mathcal{E}(\mathcal{C}) = |\mathcal{C}|^2 P_\textrm{indist} + (1 -
  |\mathcal{C}|^2) P_\textrm{dist}$, as can be confirmed using
  Eqs.~(\ref{eq:distinguishableFinalState})--(\ref{eq:distinguishableFinalState2pt2}).
\end{thebibliography}


\begin{figure*}[h]
	\includegraphics[width=\linewidth]{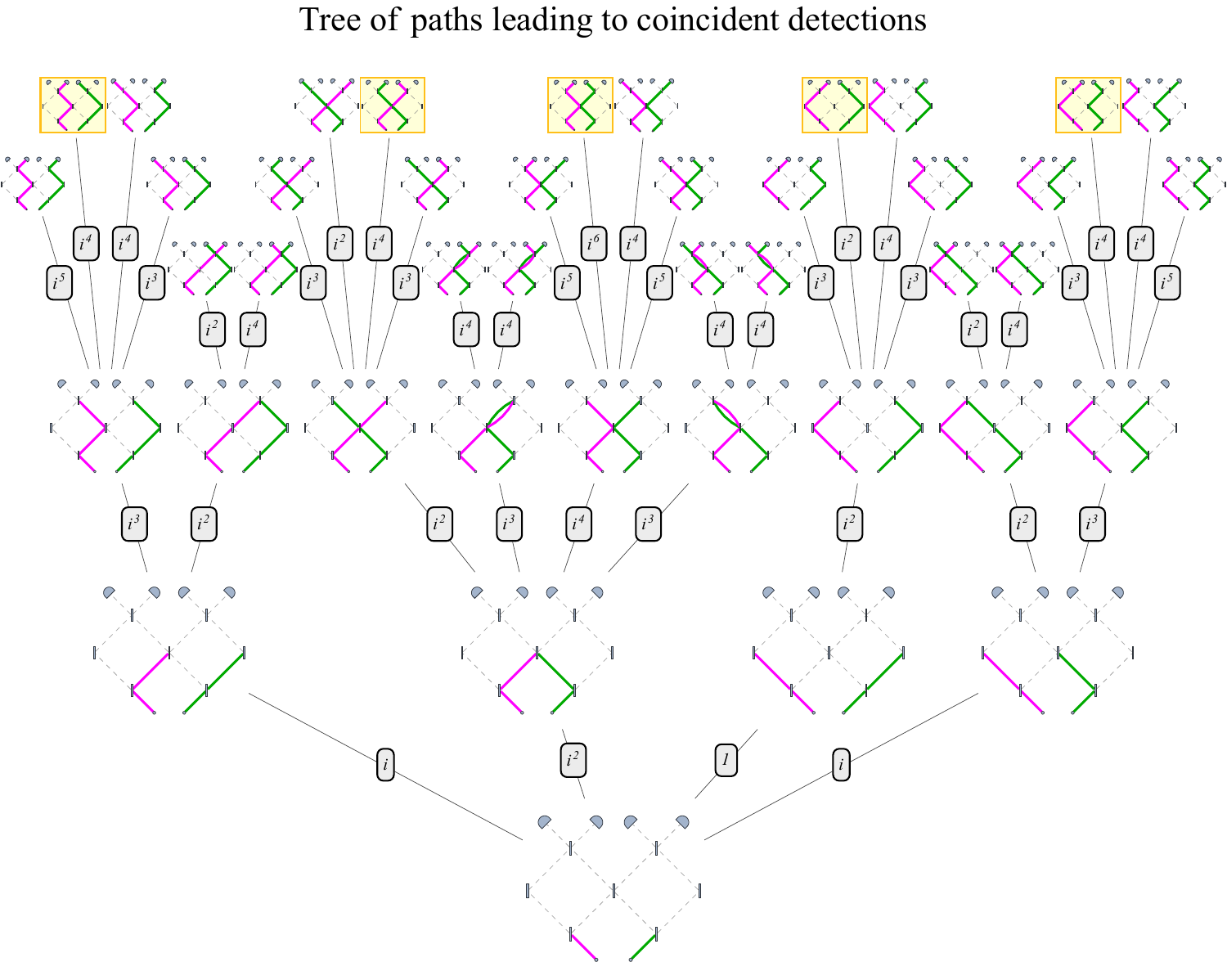}
	\caption{
		Supplemental figure.
		Visualizing all paths the photons can take through the setup that lead to a coincident detection event and counting the phase acquired (i.e., same detector two-photon detection and loss events are not shown). The five amplitudes corresponding to a D$_2$D$_3$ coincidence are highlighted in yellow. 
	}
\end{figure*}

\end{document}